\documentclass[preprint]{aastex}
\usepackage{graphicx}
\usepackage{natbib}
\begin{document}

\title{More Kronoseismology with Saturn's rings}
\author{M.M. Hedman$^a$ and P.D. Nicholson$^b$}
\affil{$^a$Department of Physics, University of Idaho, Moscow ID 83844-0903\\
$^b$Department of Astronomy, Cornell University, Ithaca NY 14850}
\maketitle

{\bf ABSTRACT:}  In a previous paper (Hedman and Nicholson 2013), we developed tools which allowed us to confirm that several of the waves in Saturn's rings were likely generated by resonances with fundamental sectoral normal modes inside Saturn itself. Here we use these same tools to examine eight additional waves that are probably generated by structures inside the planet. One of these waves appears to be generated by a resonance with a fundamental sectoral normal mode  with azimuthal  harmonic number $m=10$. If this  attribution is correct, then the $m=10$ mode must have a larger amplitude than the modes with  $m=5-9$, since the latter do not appear to generate strong waves. We also identify five waves with pattern speeds between 807$^\circ$/day and $834^\circ$/day. Since these pattern speeds are close to the planet's rotation rate, they probably are due to persistent gravitational anomalies within the planet. These waves are all found in regions of enhanced optical depth known as plateaux, but surprisingly the surface mass densities they yield are comparable to the surface mass densities of the background C ring.  Finally, one wave appears to be a one-armed spiral pattern whose rotation rate suggests it is generated by a resonance with a structure inside Saturn, but the nature of this perturbing structure remains unclear. Strangely, the resonant radius for this wave seems to be drifting inwards at an average rate of 0.8 km/year over the last thirty years, implying that the relevant planetary oscillation frequency has been steadily increasing.

\section{Introduction}

Saturn's rings exist in a complex dynamical environment, with various forces perturbing the ring-particles' orbits and forming a multitude of patterns and structures. While the gravitational tugs from Saturn's various moons have long been known to sculpt the rings, asymmetries in the planet's own gravitational field can also produce distinctive patterns in the inner part of the ring system. In particular, \citet{HN13} provided evidence that several spiral density waves in Saturn's C ring observed in Voyager and Cassini occultation data \citep{Rosen91, Colwell09, Baillie11} have the right pattern speeds and symmetry properties to be generated by low-order normal mode oscillations within Saturn, as predicted by \citet{Marley90} and \citet{MarleyPorco93}. Analyses of these patterns yielded precise estimates of the rotation periods of these modes, which should provide novel insights into the planet's deep interior. However, these measurements also revealed that there were multiple patterns with the same basic symmetry but slightly different pattern speeds. Such fine splitting of the normal modes was not predicted and might suggest that current models of Saturn's internal structure are incomplete \citep{Fuller14, Marley14}.

The analysis performed by \citet{HN13} focused on six density waves which had the strongest opacity variations and the longest wavelengths. These waves were the easiest ones to study and measure, and indeed we obtained extremely precise and robust estimates of these patterns' rotation rates around the planet. However, many more wave-like patterns exist in the C ring \citep{Baillie11}, and these same techniques could potentially be applied to these features. In this work, we estimate the pattern speeds and symmetry properties of several additional waves that provide further information about the planet's internal structure. 

Section~\ref{methods} provides a brief summary of our methods for analyzing unidentified density waves in occultation data, while Section~\ref{select} lists the known C-ring waves that could be investigated with these techniques. Section~\ref{W83.63} focuses on a wave that appears to provide another example of a feature generated by a resonance with  a low-order normal mode oscillation within the planet. This feature raises interesting questions about the amplitude spectra of these normal modes. Another, weaker wave in the same region that may be generated by a planetary normal mode is considered in Section~\ref{W81.02}. Unfortunately, the signal-to-noise of this feature turns out to be too low to yield a definitive identification at this time. Section~\ref{m3waves} considers several patterns that do not appear to be driven by normal-mode oscillations, but instead could be caused by gravitational anomalies moving around the planet at close to Saturn's rotation rate. Besides providing new information about the planet's internal structure, these waves reveal surprising trends in the C-ring's surface mass density. Section~\ref{W85.67} describes a curious wave that appears to be propagating in the wrong direction given its pattern speed, and has been moving slowly through the rings over the last 30 years.  Finally, we summarize the conclusions of this investigation.

\section{Methods}
\label{methods}

As in \citet{HN13}, we use wavelet-based techniques in order to ascertain the symmetry properties and pattern speeds of various density-wave patterns observed in multiple occultation profiles obtained by the Visual and Infrared Mapping Spectrometer (VIMS) onboard the Cassini spacecraft. For the sake of completeness, we will provide a summary of our analysis methods, and refer the reader to \citet{HN13}  for more details on these techniques.

During an occultation, VIMS measures the apparent brightness of a star as a function of time while the star passes behind the rings. Since VIMS has a highly linear response curve \citep{Brown04}, these measurements of the star's brightness can be readily transformed into estimates of the ring's opacity at the location where the starlight passes through the rings.  This opacity can be expressed in terms of the transmission $T$, the optical depth along the line of sight $\tau=-\ln(T)$ or the normal optical depth $\tau_n=\tau\sin |B|$ (where $B$ is the angle between the line of sight to the star and the ringplane).  The location where the light passed through the ring during each measurement is estimated using a combination of the precise time stamps associated with the occultation data, the reconstructed trajectory of the spacecraft (from the NAIF website\footnote{\tt http://naif.jpl.nasa.gov/pub/naif/CASSINI/}) and the location of the star on the sky (derived from the Hipparcos catalog\footnote{{\tt http://hesarc.gsfc.nasa.gov/W3Browse/hipparcos.html}\citep{Perryman97}}, and corrected for proper motion and parallax at Saturn). Each occultation therefore yields a profile of the ring's opacity as a function of ring radius, longitude and time.  Based on the positions of sharp ring edges in these profiles, we find that our reconstruction of the occultation geometry is accurate to within several hundred meters. Furthermore, a global analysis of all the occultation data by R. French provides slight trajectory corrections. These corrections yield radial position estimates that are good to within 200 meters, which is more than sufficient for the purposes of this analysis.

Within each profile, density waves appear as quasi-periodic variations in the ring's opacity with wavelengths that systematically vary with radius (see Figures~\ref{W83.63prof}, ~\ref{m3prof} and~\ref{W85.67prof}). However, the positions of individual peaks and troughs vary from occultation to occultation because the entire wave is actually a tightly wound spiral pattern, and different occultations cut through this pattern at different locations and times. Two parameters characterize these azimuthal variations: an integer $m$ that determines the number of arms in the spiral pattern, and the speed at which the entire pattern rotates around the planet. If the spiral pattern is generated by a first-order Lindblad resonance, then this pattern speed $\Omega_p$ is given by the following expression:
\begin{equation}
\Omega_p=\frac{(m-1)n+\dot{\varpi}}{m}
\end{equation}
where $n$ is the local orbital mean motion, and $\dot{\varpi}$ is the local apsidal precession rate. Note that $\Omega_p$ can be slower or faster than $n$, depending on whether $m$ is positive or negative. Waves with $m>0$ and $\Omega_p<n$ are associated  with inner Lindblad resonances and are expected to propagate outwards, while those with $m<0$ and $\Omega_p>n$ are associated with outer Lindblad resonances and should propagate inwards. Both types of patterns have $|m|$ arms, so the azimuthal location of a given cut through the pattern can be quantified in terms of a phase parameter
\begin{equation}
\phi=|m|(\lambda-\Omega_p t) 
\end{equation}
where $\lambda$ and $t$ are the inertial longitude and time of the occultation. So long as the amplitude of the wave is small, the optical depth variations associated with the wave can be written as 
\begin{equation}
\tau(r) \simeq \tau_0 + \Delta\tau(r)\cos(\phi+\phi_r(r)),
\label{eqtau}
\end{equation}
where $\Delta\tau(r)$ is a slowly varying amplitude and $\phi_r(r)$ is a smoothly increasing function of $r$ that determines the local radial wavelength of the pattern. One can therefore regard $\phi$ as a measure of where the peaks and troughs of the profile are found in a given profile obtained at a particular longitude and time. (If we choose a suitable radius $r_0$ in the wave where $\phi_r=0$, then  $\phi=0$ corresponds to a profile where there is a opacity maximum at $r_0$, while $\phi=180^\circ$ corresponds to a profile with a minimum in opacity at the same $r_0$.) 

In practice, accurately measuring the absolute value of the phase $\phi$ for an unidentified wave is challenging because the radial wavelength of the wave varies smoothly with radius (i.e. $d\phi_r/dr$ is not a constant), so the phase estimate at a particular radius for a given occultation profile is sensitive to how well these trends can be modeled. By contrast, the phase difference between two profiles $\delta \phi$ is easier to determine because this difference should be a constant across the entire wave.  We therefore compute $\delta \phi$ for each pair of occultations using wavelet transforms. Again, we will simply outline our procedures here, and refer the interested reader to \citet{HN13} for the relevant details. First, we interpolate the data from each occultation onto a regular array of radii and compute a wavelet transform using the standard IDL routine {\tt wavelet} \citep{TC98}. Next,  we compute the wavelet power and weighted average wavelet phase $\phi+\phi_r$ for each profile as a function of radius in a specified spatial wavelength band. In \citet{HN13} the wavelength range used was from 0.1 km to 5 km, but here we will use different ranges for different waves.  Next, we compute $\delta \phi$ between two cuts as the weighted average difference of the two phase profiles, using the average power in the two profiles as the weighting function.

As before, we visually inspect the occultation profiles and only consider observations where the wave is clearly resolved and data gaps do not corrupt multiple wave cycles. We also automatically exclude $\delta\phi$ estimates from any pair of occultations where the average of the two power profiles is never more than 0.9 times the normalized peak signal in either individual profile (which indicates the peak signal is being measured in disjoint regions) or if the standard deviation of the phase differences to be averaged together exceeds 20$^\circ$ (which indicates the phase difference is not being accurately measured). Depending on the analysis, we also exclude pairs with time separations longer than either 300 days or 1000 days, simply to reduce the effects of aliasing.

For any given trial values of $m$ and pattern speed $\Omega_P$, we can compute the expected phase difference between each pair of profiles:
\begin{equation}
\delta \phi_{pred}=|m|(\delta \lambda-\Omega_p \delta t)
\label{deltaphi}
\end{equation}
where $\delta \lambda$ and $\delta t$ are the difference in the observed  inertial longitudes and observation times for the two occultations. We then compare these predicted values to the observed values of $\delta \phi$ to ascertain which $m$ and $\Omega_p$ are most consistent with the observations. In practice, we examine the $rms$ dispersion of the  ``phase difference residuals'' (i.e. the observed values of $\delta \phi$ minus the predicted values). For most of the waves considered here, we find that for one particular value of $m$, there is a pronounced minimum in the $rms$ residuals near the predicted pattern speed. This minimum thus provides a fairly unique identification of the symmetry properties and pattern speed of the relevant wave.

\section{Selecting Features to Examine}
\label{select}

\begin{table}
\caption{Summary of wave-like C-ring features}
\label{wavelist}
\resizebox{6in}{!}{\begin{tabular}{|cccc|}\hline
Baillie {\it et al.} Feature & Location & Identification & Comment \\ \hline
1  & 74666 km & & Not examined, too few VIMS occs with sufficient resolution \\
2  & 74892 km & Mimas 4:1 ILR & See Footnote 3 \\
3  & 74923 km & & Not examined, too few VIMS occs with sufficient resolution \\
4  & 74939 km & Rosen waveb & Not examined, too few VIMS occs with sufficient resolution \\
5  & 76022 km & & Not examined, too few VIMS occs with sufficient resolution \\
6  & 76234 km & & Not examined, too few VIMS occs with sufficient resolution \\
9  & 76435 km &  & Not examined, too few VIMS occs with sufficient resolution \\
10 & 76539 km & & Not examined, too few VIMS occs with sufficient resolution \\
11 & 76729 km & & Not examined, too few VIMS occs with sufficient resolution \\
12  & 77511 km & Titan -1:0 IVR & See Nicholson et al. (in prep)  \\
13 & 80988 km &  W80.98 m=-4 pattern & Examined in Hedman and Nicholson (2013) \\
14 & 81018 km &  W81.02 (m=-11? -5?) &  Examined in Section~\ref{W81.02} \\
15 & 82010 km & W82.01 m=-3 pattern & Examined in Hedman and Nicholson (2013) \\
16 & 82061 km & W82.06 m=-3 pattern & Examined in Hedman and Nicholson (2013) \\
17 & 82209 km & W82.21 m=-3 pattern & Examined in Hedman and Nicholson (2013) \\
18 & 83633 km & W83.63 m=-10 pattern & Examined in Section~\ref{W83.63} \\
19 & 84644 km & W84.64 m=-2 pattern & Examined in Hedman and Nicholson (2013) \\
20 & 84814 km & W84.82 m=+3 pattern & Examined in Section~\ref{m3waves} \\
21 & 84857 km & W84.86 m=+3 pattern & Examined in Section~\ref{m3waves} \\
22 & 85105 km & Pan 2:1 ILR & Not examined here, too low signal-to-noise \\
23 & 85450 km & Rosen wave j & Not examined here, too low signal-to-noise \\
24 & 85473 km & & Not examined here, too low signal-to-noise \\
25 & 85514 km & & Not examined here, too low signal-to-noise \\
26 & 85523 km & & Not examined here, too low signal-to-noise \\
27 & 85677 km & W85.67 m=-1 pattern & Examined in Section~\ref{W85.67} \\
28 & 86400 km & W86.40 m=+3 pattern & Examined in Section~\ref{m3waves} \\
29 & 86576 km & W86.58 m=+3 pattern & Examined in Section~\ref{m3waves}\\
31 & 86590 km & W86.59 m=+3 pattern & Examined in Section~\ref{m3waves} \\
32 & 87189 km & W87.19 m=-2 pattern & Examined in Hedman and Nicholson (2013) \\
33 & 87647 km & Atlas 2:1 ILR & Not examined here, too low signal-to-noise \\ 
34 & 88713 km & Prometheus 2:1 ILR & Not examined here, on a complex ringlet \\
35 & 88736 km & m=+1  &  See Footnote 4 \\
36 & 89883 km & Mimas 6:2 ILR & Examined in Section~\ref{m3waves} \\
37 & 89894 km & Pandora 4:2 ILR & Examined in Section~\ref{m3waves} \\ 
38 & 90156 km & Pandora 2:1 ILR (?) & Not examined here, on complex ringlet \\
39 & 90198 km & Mimas 3:1 ILR & Not examined here, on complex ringlet \\
40 & 90279 km & & Not examined, too low signal-to-noise \\
\hline \end{tabular}}
\end{table}

\begin{figure}
\resizebox{6in}{!}{\includegraphics{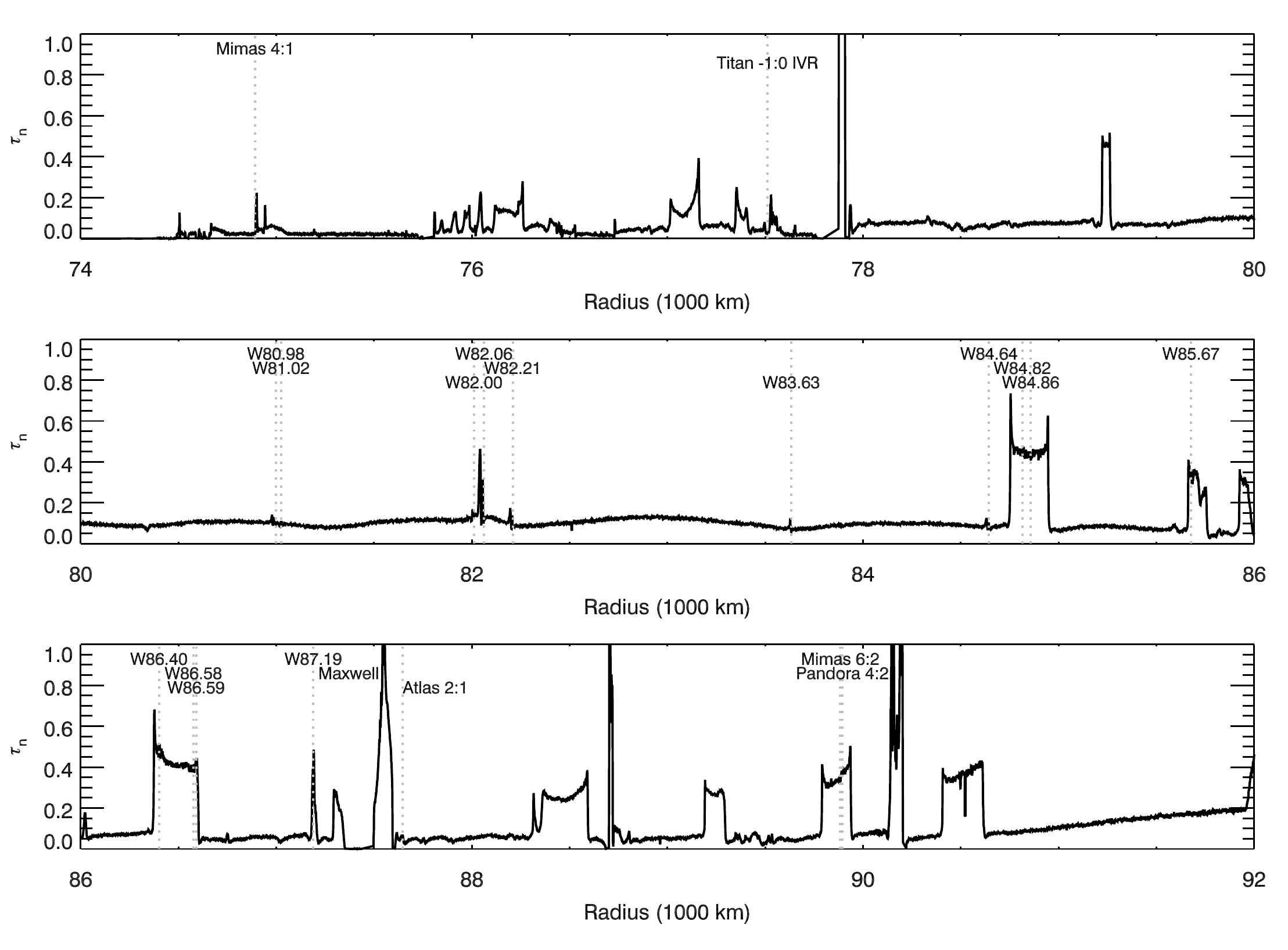}}
\caption{Optical depth profile of the C ring, derived from the $\gamma$ Crucis occultation obtained during Cassini orbit (``Rev") 89. The profile has been downsampled for display purposes, and the locations of the various waves discussed in the text are marked.}
\label{overview}
\end{figure}

\begin{figure}
\centerline{\resizebox{5in}{!}{\includegraphics{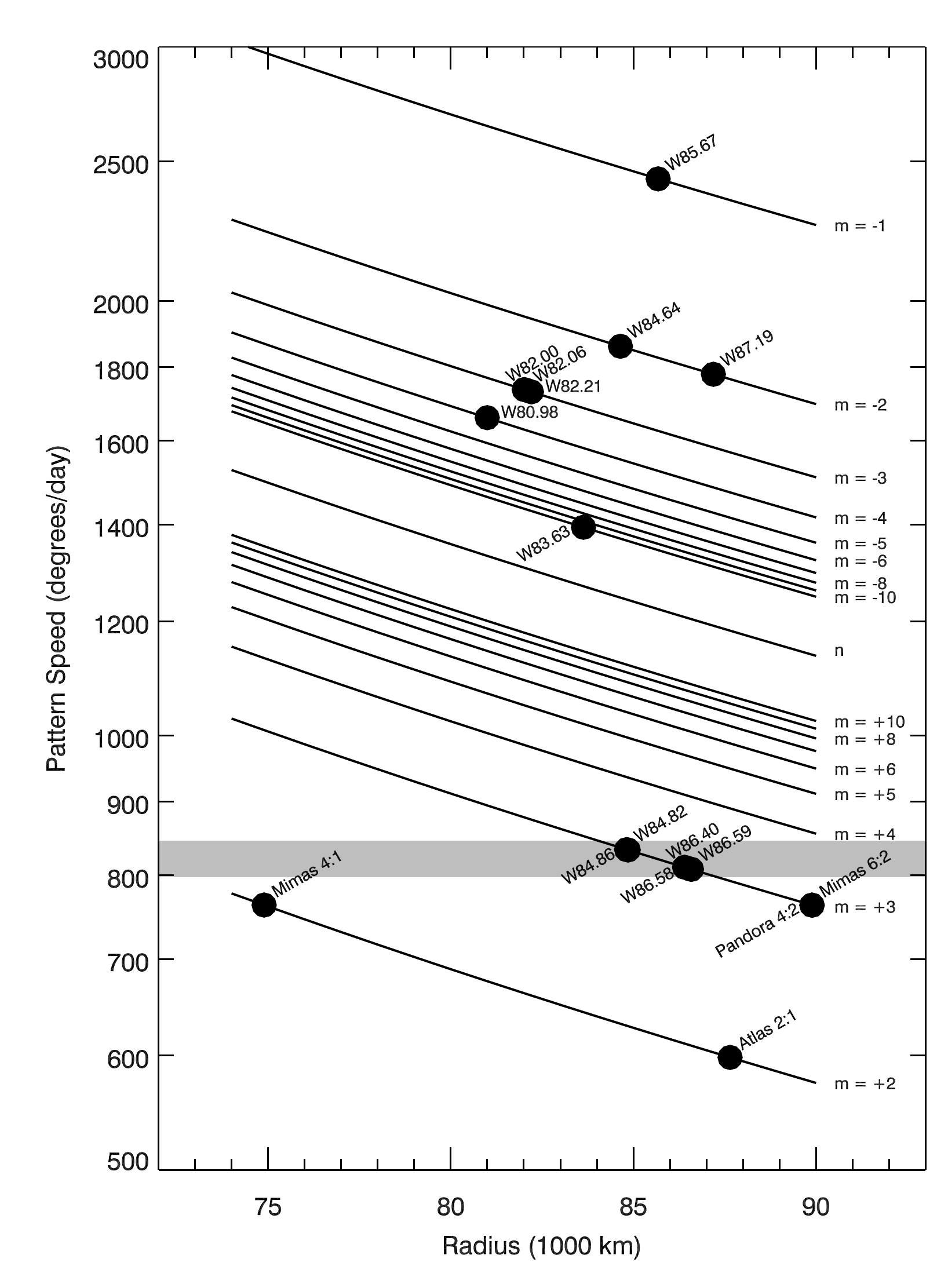}}}
\caption{The locations and pattern speeds of various waves in Saturn's C ring. This plot shows the radius (distance from Saturn center) and pattern speeds for many of the waves identified by \citet{Baillie11}, \citet{HN13} and this work (no data point is given for wave W81.02 because the identification of this wave is still uncertain). The series of diagonal lines correspond to the expected pattern speeds for first-order Lindblad resonances with different values of $m$. The lines with negative $m$ values correspond to outer Lindblad resonances, while the lines with positive $m$ values correspond to inner Lindblad resonances. The line marked as $n$ is where the pattern speed would equal the local orbital rate. The grey shaded band corresponds to the range of rotation rates observed in Saturn's visible winds and magnetospheric radio emissions (see Figure~\ref{satper}). }
\label{reslocs}
\end{figure}

We can only extract sensible phase estimates if the occultation profiles have sufficient resolution and signal-to-noise, so we cannot use the techniques described above if the wave's wavelength or amplitude is too small. We therefore conducted a survey of the roughly 40 wave-like features in the C ring identified by \citet{Baillie11} in order to ascertain which patterns would be the best ones to investigate. Table~\ref{wavelist} summarizes the results of this survey. 

There are several sets of waves we are unable to analyze using the currently-available VIMS data and the routines described above. More specifically, the features interior to 77,000 km all have rather short wavelengths and we do not have enough VIMS occultations with sufficient resolution to determine the relevant phase differences.\footnote{We did have sufficient VIMS occultations to confirm that the wave near the Mimas 4:1 inner Lindblad resonance was an $m=2$ pattern with the correct pattern speed for this resonance.} Investigations of these structures will therefore likely require higher-resolution ($<$ 200 m) data from the UVIS and RSS experiments. We chose  not to analyze any of the five waves \citet{Baillie11} found between 85,000 and 85,500 km,  the Atlas 2:1 wave at 87,647 km or the wave at 90,279 km  because these features are very weak and could not be clearly discerned in individual VIMS occultation profiles. (Indeed, \citet{Baillie11} were only able to detect some of these weaker waves by co-adding data from multiple occultations.)   Finally, we have elected not to consider the Titan -1:0 bending wave at 77,520 km, or any of the wave-like patterns found within or close to the C-ring's various dense ringlets. These structures are in complex dynamical environments that will likely require modified processing algorithms that we plan to develop in a future work.\footnote{We did find one wave-like structure at 88,736  km (designated feature 35 by Baillie {\it et al.} 2011), which could be analyzed using the tools described above. The phase differences derived from that wave indicate that it is an $m=+1$ spiral pattern rotating around the planet at around 14 $^\circ$/day. This pattern speed is much slower any predicted resonance with Saturn and instead may indicate some connection with the adjacent Bond ringlet. We will therefore postpone a detailed investigation of this feature to a paper focusing on the waves associated with dense ringlets.} 

In the end, we found sixteen waves that we could productively investigate at this time. The locations of all these features are marked in Figure~\ref{overview}, while Table~\ref{wavelist} and Figure~\ref{reslocs} summarize our conclusions regarding the pattern speeds and symmetry properties for many of these waves. Our analyses indicate that these features can be divided into three broad classes. First, we have the waves that appear to be generated by sectoral normal modes inside the planet. These include the six waves examined in our previous paper, along with two features at 81,018 km and 83,633 km that are in the same general part of the rings.  Second, we have seven fairly weak waves within features known as ``plateaux'' in the outer part of the C ring. Two of these features were already identified with the 6:2 Mimas and 4:2 Pandora resonances \citep{Baillie11}, while the other five (found inside plateaux P5 and P7) appear to be $m=+3$ patterns with pattern speeds close to the planet's rotation rate. Finally, there is a very unusual wave in plateau P6 around 85,670 km which actually appears to have moved steadily inwards over the last 30 years.

Tables~\ref{obstab} and~\ref{obstab3} provide lists of the VIMS occultations that  were considered in the analysis of each of these waves. Lists of phase difference estimates derived from these profiles are available in an online data supplement, a sample of which is shown in Table~\ref{phasetab}. For a summary of the results of this investigation, including the inferred $m$-numbers and pattern speeds of the relevant waves, see Table~\ref{sumtab}. Note that this analysis includes several occultations that were not part of our earlier study. Re-analyzing the original six waves with this larger data set yielded essentially the same results as our previous analysis (compare the relevant entries in Table~\ref{sumtab} with Table 4 of Hedman and Nicholson 2013). 

\begin{table}
\caption{Observed times (in seconds of ephemeris time, measured from the J2000 epoch) and inertial longitudes (measured relative to the longitude of ascending node on the J2000 equator) for the occultation cuts through the various waves associated with the outer Lindblad resonances. Blank entries indicate that occultation did not provide a suitable profile for that wave.}
\label{obstab}
\resizebox{!}{4in}{\begin{tabular}{|c|c|c| c|c|c|c|c|c|c|c|c|}\hline
Star & Rev & & W80.98 & W81.02 & W82.00 & W82.06 & W82.21 & W83.63 & W84.64 & W85.67 & W87.19 \\ \hline\hline
R Hya&036&i&      220948260.&      220948254.&      220948093.&      220948085.&      220948061.&      220947834.&      220947678.&      220947519.&      220947295.\\
&&&         174.072$^\circ$&         174.109$^\circ$&         175.060$^\circ$&         175.108$^\circ$&         175.249$^\circ$&         176.544$^\circ$&         177.411$^\circ$&         178.273$^\circ$&         179.453$^\circ$\\ \hline
$\alpha$ Aur&041&i&      227949007.&&      227948789.&      227948779.&      227948748.&      227948458.&      227948260.&      227948061.&      227947783.\\
&&&         343.813$^\circ$&&         345.077$^\circ$&         345.137$^\circ$&         345.316$^\circ$&         346.944$^\circ$&         348.019$^\circ$&         349.077$^\circ$&         350.513$^\circ$\\ \hline
$\gamma$ Cru&071&i&      266193414.&      266193408.&      266193267.&      266193260.&      266193238.&&      266192888.&      266192738.&      266192521.\\
&&&         183.104$^\circ$&         183.109$^\circ$&         183.241$^\circ$&         183.248$^\circ$&         183.267$^\circ$&&         183.579$^\circ$&         183.707$^\circ$&         183.886$^\circ$\\ \hline
$\gamma$ Cru&073&i&      267426088.&&      267425942.&      267425934.&      267425913.&      267425709.&      267425565.&      267425415.&      267425200.\\
&&&         182.136$^\circ$&&         182.281$^\circ$&         182.288$^\circ$&         182.309$^\circ$&         182.506$^\circ$&         182.641$^\circ$&         182.778$^\circ$&         182.969$^\circ$\\ \hline
$\gamma$ Cru&077&i&      269858216.&&      269858071.&      269858064.&      269858043.&&      269857698.&      269857550.&      269857336.\\
&&&         181.086$^\circ$&&         181.240$^\circ$&         181.247$^\circ$&         181.269$^\circ$&&         181.619$^\circ$&         181.763$^\circ$&         181.965$^\circ$\\ \hline
$\gamma$ Cru&078&i&      270466692.&&      270466548.&      270466541.&      270466520.&      270466318.&      270466175.&      270466028.&      270465815.\\
&&&         180.860$^\circ$&&         181.015$^\circ$&         181.023$^\circ$&         181.045$^\circ$&         181.255$^\circ$&         181.398$^\circ$&         181.544$^\circ$&         181.747$^\circ$\\ \hline
$\beta$ Gru&078&i&&&&&&&      270512794.&      270512542.&      270512284.\\
&&&&&&&&&         302.774$^\circ$&         298.579$^\circ$&         294.401$^\circ$\\ \hline
$\gamma$ Cru&079&i&      271045522.&&      271045367.&      271045359.&      271045337.&&      271044967.&      271044809.&      271044581.\\
&&&         179.175$^\circ$&&         179.354$^\circ$&         179.363$^\circ$&         179.389$^\circ$&&         179.796$^\circ$&         179.964$^\circ$&         180.198$^\circ$\\ \hline
RS Cnc&080&i&      271872473.&      271872458.&      271872088.&      271872070.&      271872018.&      271871557.&      271871263.&      271870979.&\\
&&&          90.240$^\circ$&          90.145$^\circ$&          87.858$^\circ$&          87.749$^\circ$&          87.431$^\circ$&          84.679$^\circ$&          82.976$^\circ$&          81.365$^\circ$&\\ \hline
RS Cnc&080&e&      271877226.&      271877241.&      271877611.&      271877629.&      271877681.&      271878142.&      271878435.&      271878720.&\\
&&&         121.515$^\circ$&         121.610$^\circ$&         123.897$^\circ$&         124.005$^\circ$&         124.324$^\circ$&         127.075$^\circ$&         128.778$^\circ$&         130.388$^\circ$&\\ \hline
$\gamma$ Cru&081&i&      272320388.&      272320382.&      272320233.&      272320226.&      272320203.&      272319987.&      272319834.&      272319676.&      272319448.\\
&&&         178.322$^\circ$&         178.329$^\circ$&         178.510$^\circ$&         178.519$^\circ$&         178.546$^\circ$&         178.800$^\circ$&         178.974$^\circ$&         179.151$^\circ$&         179.397$^\circ$\\ \hline
$\gamma$ Cru&082&i&      272956171.&      272956166.&      272956016.&      272956009.&      272955986.&&      272955616.&      272955457.&      272955229.\\
&&&         177.862$^\circ$&         177.869$^\circ$&         178.056$^\circ$&         178.065$^\circ$&         178.093$^\circ$&&         178.533$^\circ$&         178.714$^\circ$&         178.967$^\circ$\\ \hline
RS Cnc&085&i&&&      275057262.&      275057224.&      275057119.&      275056348.&      275055930.&      275055553.&      275055073.\\
&&&&&          97.236$^\circ$&          97.000$^\circ$&          96.337$^\circ$&          91.565$^\circ$&          89.056$^\circ$&          86.845$^\circ$&          84.114$^\circ$\\ \hline
RS Cnc&085&e&&&      275059898.&      275059935.&      275060040.&      275060811.&      275061229.&      275061606.&      275062086.\\
&&&&&         114.136$^\circ$&         114.371$^\circ$&         115.034$^\circ$&         119.806$^\circ$&         122.315$^\circ$&         124.525$^\circ$&         127.256$^\circ$\\ \hline
$\gamma$ Cru&086&i&      275503697.&      275503692.&      275503542.&      275503535.&      275503512.&&&      275502984.&      275502756.\\
&&&         176.829$^\circ$&         176.837$^\circ$&         177.033$^\circ$&         177.043$^\circ$&         177.073$^\circ$&&&         177.728$^\circ$&         177.995$^\circ$\\ \hline
RS Cnc&087&i&&&&&&      276330567.&      276329996.&      276329532.&      276328974.\\
&&&&&&&&          96.181$^\circ$&          92.695$^\circ$&          89.925$^\circ$&          86.701$^\circ$\\ \hline
RS Cnc&087&e&&&&&&      276333572.&      276334143.&      276334608.&      276335165.\\
&&&&&&&&         115.078$^\circ$&         118.564$^\circ$&         121.334$^\circ$&         124.557$^\circ$\\ \hline
$\gamma$ Cru&089&i&      277408751.&      277408745.&      277408596.&      277408589.&      277408566.&&      277408198.&      277408040.&      277407813.\\
&&&         176.576$^\circ$&         176.584$^\circ$&         176.781$^\circ$&         176.791$^\circ$&         176.821$^\circ$&&         177.288$^\circ$&         177.480$^\circ$&         177.749$^\circ$\\ \hline
$\gamma$ Cru&093&i&      280045204.&      280045198.&      280045028.&      280045020.&      280044994.&&      280044575.&      280044395.&      280044136.\\
&&&         208.249$^\circ$&         208.242$^\circ$&         208.061$^\circ$&         208.052$^\circ$&         208.024$^\circ$&&         207.597$^\circ$&         207.421$^\circ$&         207.175$^\circ$\\ \hline
$\gamma$ Cru&094&i&      280681410.&&      280681250.&      280681242.&      280681218.&&      280680835.&      280680670.&      280680433.\\
&&&         191.683$^\circ$&&         191.696$^\circ$&         191.697$^\circ$&         191.699$^\circ$&&         191.728$^\circ$&         191.741$^\circ$&         191.758$^\circ$\\ \hline
$\gamma$ Cru&096&i&      282014259.&      282014253.&      282014111.&      282014104.&      282014083.&      282013876.&      282013731.&      282013580.&      282013362.\\
&&&         185.190$^\circ$&         185.193$^\circ$&         185.280$^\circ$&         185.285$^\circ$&         185.298$^\circ$&         185.420$^\circ$&         185.504$^\circ$&         185.589$^\circ$&         185.708$^\circ$\\ \hline
$\gamma$ Cru&100&i&      285034037.&&      285033857.&      285033848.&      285033822.&&      285033397.&      285033216.&      285032956.\\
&&&         224.282$^\circ$&&         223.835$^\circ$&         223.814$^\circ$&         223.749$^\circ$&&         222.739$^\circ$&         222.326$^\circ$&         221.750$^\circ$\\ \hline
$\gamma$ Cru&101&i&      285861190.&      285861183.&      285861011.&      285861002.&      285860975.&&      285860551.&      285860370.&      285860110.\\
&&&         224.289$^\circ$&         224.272$^\circ$&         223.842$^\circ$&         223.820$^\circ$&         223.755$^\circ$&&         222.744$^\circ$&         222.331$^\circ$&         221.755$^\circ$\\ \hline
$\gamma$ Cru&102&i&      286686360.&      286686354.&      286686182.&      286686173.&      286686147.&&      286685724.&      286685544.&      286685285.\\
&&&         223.942$^\circ$&         223.926$^\circ$&         223.500$^\circ$&         223.479$^\circ$&         223.415$^\circ$&&         222.416$^\circ$&         222.007$^\circ$&         221.438$^\circ$\\ \hline
$\beta$ Peg&104&i&      288914432.&      288914429.&      288914336.&      288914332.&      288914318.&&      288914090.&      288913993.&\\
&&&         342.574$^\circ$&         342.591$^\circ$&         343.021$^\circ$&         343.042$^\circ$&         343.107$^\circ$&&         344.115$^\circ$&         344.528$^\circ$&\\ \hline
R Cas&106&i&      291039691.&      291039659.&      291038969.&      291038939.&      291038853.&&      291037728.&      291037338.&      291036830.\\
&&&          90.705$^\circ$&          90.524$^\circ$&          86.728$^\circ$&          86.566$^\circ$&          86.097$^\circ$&&          80.176$^\circ$&          78.210$^\circ$&          75.723$^\circ$\\ \hline
$\alpha$ Sco&115&i&      302022977.&      302022964.&      302022637.&      302022621.&      302022571.&&&      302021428.&      302020939.\\
&&&         157.895$^\circ$&         157.914$^\circ$&         158.409$^\circ$&         158.434$^\circ$&         158.508$^\circ$&&&         160.139$^\circ$&         160.797$^\circ$\\ \hline
$\beta$ Peg&170&e&      397973362.&      397973371.&&      397973608.&      397973643.&      397973970.&&      397974439.&      397974782.\\
&&&          78.465$^\circ$&          78.478$^\circ$&&          78.837$^\circ$&          78.888$^\circ$&          79.366$^\circ$&&          80.022$^\circ$&          80.482$^\circ$\\ \hline
$\beta$ Peg&172&i&      401620518.&      401620509.&      401620280.&      401620269.&      401620234.&      401619902.&      401619669.&      401619428.&      401619082.\\
&&&         312.326$^\circ$&         312.312$^\circ$&         311.950$^\circ$&         311.932$^\circ$&         311.877$^\circ$&         311.371$^\circ$&         311.025$^\circ$&         310.676$^\circ$&         310.189$^\circ$\\ \hline
$\lambda$ Vel&173&i&&&&&&&&      403833306.&\\
&&&&&&&&&&         147.873$^\circ$&\\ \hline
$\alpha$ Lyr&175&i&&&&&&&&      406601709.&\\
&&&&&&&&&&         268.044$^\circ$&\\ \hline
W Hya&179&i&      411902978.&      411902973.&      411902841.&      411902834.&      411902814.&&&      411902358.&\\
&&&         146.520$^\circ$&         146.494$^\circ$&         145.824$^\circ$&         145.791$^\circ$&         145.690$^\circ$&&&         143.508$^\circ$&\\ \hline
W Hya&180&i&&&&      413052249.&      413052229.&      413052038.&      413051906.&      413051769.&      413051575.\\
&&&&&&         146.312$^\circ$&         146.210$^\circ$&         145.265$^\circ$&         144.625$^\circ$&         143.984$^\circ$&         143.097$^\circ$\\ \hline
W Hya&181&i&      414201633.&      414201628.&      414201494.&&      414201467.&      414201277.&      414201144.&      414201007.&\\
&&&         147.135$^\circ$&         147.108$^\circ$&         146.422$^\circ$&&         146.285$^\circ$&         145.336$^\circ$&         144.694$^\circ$&         144.051$^\circ$&\\ \hline
R Cas&185&i&      418067466.&&&      418067199.&      418067162.&&&      418066320.&      418065965.\\
&&&         331.261$^\circ$&&&         330.441$^\circ$&         330.329$^\circ$&&&         327.890$^\circ$&         326.923$^\circ$\\ \hline
$\mu$ Cep&185&e&&      418015692.&      418016089.&      418016110.&      418016169.&      418016725.&      418017105.&      418017491.&      418018032.\\
&&&&          44.302$^\circ$&          45.391$^\circ$&          45.445$^\circ$&          45.606$^\circ$&          47.076$^\circ$&          48.053$^\circ$&          49.019$^\circ$&          50.337$^\circ$\\ \hline
$\gamma$ Cru&187&i&      419919773.&      419919766.&      419919584.&      419919575.&      419919547.&      419919290.&      419919112.&      419918931.&      419918675.\\
&&&         151.851$^\circ$&         151.815$^\circ$&         150.890$^\circ$&         150.844$^\circ$&         150.707$^\circ$&         149.443$^\circ$&         148.597$^\circ$&         147.754$^\circ$&         146.597$^\circ$\\ \hline
$\gamma$ Cru&187&e&      419930501.&      419930508.&      419930689.&      419930699.&      419930726.&      419930983.&      419931160.&      419931341.&      419931596.\\
&&&         225.432$^\circ$&         225.468$^\circ$&         226.391$^\circ$&         226.438$^\circ$&         226.575$^\circ$&         227.836$^\circ$&         228.681$^\circ$&         229.522$^\circ$&         230.676$^\circ$\\ \hline
R Cas&191&i&      423133316.&      423133309.&&      423133127.&      423133100.&&      423132671.&      423132487.&\\
&&&         296.924$^\circ$&         296.917$^\circ$&&         296.706$^\circ$&         296.676$^\circ$&&         296.202$^\circ$&         296.007$^\circ$&\\ \hline
$\mu$ Cep&191&i&&&      423056492.&      423056481.&&&      423055942.&      423055724.&      423055410.\\
&&&&&         290.148$^\circ$&         290.141$^\circ$&&&         289.828$^\circ$&         289.707$^\circ$&         289.538$^\circ$\\ \hline
$\mu$ Cep&193&i&      425123419.&&      425123206.&      425123195.&      425123164.&&      425122656.&      425122438.&      425122124.\\
&&&         290.475$^\circ$&&         290.342$^\circ$&         290.336$^\circ$&         290.316$^\circ$&&         290.014$^\circ$&         289.890$^\circ$&         289.716$^\circ$\\ \hline
R Cas&194&e&&&&      426260018.&&&      426260474.&      426260659.&      426260924.\\
&&&&&&          84.264$^\circ$&&&          84.782$^\circ$&          84.982$^\circ$&          85.262$^\circ$\\ \hline
\end{tabular}}
\end{table}

\begin{table}
\caption{Observed times (in seconds of ephemeris time, measured from the J2000 epoch) and inertial longitudes (measured relative to the longitude of ascending node on the J2000 equator) for the occultation cuts through the various $m=+3$ waves. Blank entries indicate that occultation did not provide a suitable profile for that wave.}
\label{obstab3}
\resizebox{!}{4in}{\begin{tabular}{|c|c|c| c|c|c|c|c|c|c|c|}\hline
Star & Rev & & W84.82 & W84.86 & W86.40 (inner) & W86.40 (outer) & W86.58 & W86.59 & Mimas 6:2 & Pandora 4:2 \\ \hline\hline 
R Hya&036&i&      220947650.&      220947643.&      220947411.&      220947406.&      220947385.&      220947382.&      220946906.&      220946903.\\
&&&         177.564$^\circ$&         177.602$^\circ$&         178.851$^\circ$&         178.875$^\circ$&         178.984$^\circ$&         179.000$^\circ$&         181.407$^\circ$&         181.421$^\circ$\\ \hline
$\alpha$ Aur&041&i&      227948225.&      227948217.&      227947926.&      227947920.&      227947895.&      227947891.&      227947307.&      227947304.\\
&&&         348.207$^\circ$&         348.254$^\circ$&         349.782$^\circ$&         349.811$^\circ$&         349.944$^\circ$&         349.963$^\circ$&         352.857$^\circ$&         352.873$^\circ$\\ \hline
$\gamma$ Cru&071&i&      266192862.&      266192855.&      266192633.&      266192629.&      266192609.&      266192606.&      266192133.&      266192131.\\
&&&         183.601$^\circ$&         183.607$^\circ$&         183.794$^\circ$&         183.798$^\circ$&         183.814$^\circ$&         183.817$^\circ$&         184.193$^\circ$&         184.195$^\circ$\\ \hline
$\gamma$ Cru&073&i&      267425539.&      267425532.&      267425311.&      267425307.&      267425287.&      267425284.&      267424814.&      267424811.\\
&&&         182.665$^\circ$&         182.671$^\circ$&         182.871$^\circ$&         182.874$^\circ$&         182.892$^\circ$&         182.895$^\circ$&         183.295$^\circ$&         183.297$^\circ$\\ \hline
$\gamma$ Cru&077&i&&&      269857447.&      269857443.&      269857423.&      269857420.&&\\
&&&&&         181.862$^\circ$&         181.866$^\circ$&         181.884$^\circ$&         181.887$^\circ$&&\\ \hline
$\gamma$ Cru&078&i&      270466149.&      270466143.&      270465925.&      270465920.&      270465901.&      270465898.&&\\
&&&         181.424$^\circ$&         181.430$^\circ$&         181.643$^\circ$&         181.647$^\circ$&         181.666$^\circ$&         181.668$^\circ$&&\\ \hline
$\beta$ Gru&078&i&      270512741.&      270512729.&      270512409.&      270512404.&      270512380.&      270512377.&&\\
&&&         301.891$^\circ$&         301.686$^\circ$&         296.395$^\circ$&         296.313$^\circ$&         295.935$^\circ$&         295.882$^\circ$&&\\ \hline
$\gamma$ Cru&079&i&      271044940.&      271044933.&      271044699.&      271044694.&      271044673.&      271044670.&      271044172.&      271044169.\\
&&&         179.826$^\circ$&         179.833$^\circ$&         180.078$^\circ$&         180.083$^\circ$&         180.104$^\circ$&         180.108$^\circ$&         180.598$^\circ$&         180.601$^\circ$\\ \hline
RS Cnc&080&i&      271871212.&      271871200.&      271870791.&      271870783.&      271870748.&      271870743.&      271869972.&      271869968.\\
&&&          82.684$^\circ$&          82.612$^\circ$&          80.321$^\circ$&          80.280$^\circ$&          80.086$^\circ$&          80.058$^\circ$&          75.991$^\circ$&          75.969$^\circ$\\ \hline
RS Cnc&080&e&      271878486.&      271878499.&      271878908.&      271878915.&      271878951.&      271878956.&      271879726.&      271879730.\\
&&&         129.070$^\circ$&         129.141$^\circ$&         131.432$^\circ$&         131.473$^\circ$&         131.667$^\circ$&         131.695$^\circ$&         135.761$^\circ$&         135.783$^\circ$\\ \hline
$\gamma$ Cru&081&i&      272319806.&      272319799.&      272319566.&      272319561.&      272319540.&      272319537.&      272319039.&      272319036.\\
&&&         179.005$^\circ$&         179.013$^\circ$&         179.271$^\circ$&         179.275$^\circ$&         179.298$^\circ$&         179.302$^\circ$&         179.817$^\circ$&         179.820$^\circ$\\ \hline
$\gamma$ Cru&082&i&      272955588.&      272955581.&      272955347.&      272955342.&      272955321.&      272955318.&      272954819.&      272954816.\\
&&&         178.565$^\circ$&         178.573$^\circ$&         178.838$^\circ$&         178.843$^\circ$&         178.866$^\circ$&         178.869$^\circ$&         179.400$^\circ$&         179.403$^\circ$\\ \hline
RS Cnc&085&i&      275055861.&      275055844.&      275055313.&      275055304.&      275055260.&      275055253.&      275054331.&      275054326.\\
&&&          88.646$^\circ$&          88.546$^\circ$&          85.471$^\circ$&          85.416$^\circ$&          85.166$^\circ$&          85.130$^\circ$&          80.087$^\circ$&          80.061$^\circ$\\ \hline
RS Cnc&085&e&      275061298.&      275061315.&      275061846.&      275061855.&      275061899.&      275061905.&      275062828.&      275062833.\\
&&&         122.725$^\circ$&         122.824$^\circ$&         125.899$^\circ$&         125.953$^\circ$&         126.204$^\circ$&         126.240$^\circ$&         131.281$^\circ$&         131.308$^\circ$\\ \hline
$\gamma$ Cru&086&i&&      275503108.&      275502874.&      275502869.&      275502848.&      275502845.&      275502347.&      275502344.\\
&&&&         177.579$^\circ$&         177.858$^\circ$&         177.863$^\circ$&         177.888$^\circ$&         177.892$^\circ$&         178.451$^\circ$&         178.454$^\circ$\\ \hline
RS Cnc&087&i&      276329908.&      276329887.&      276329250.&      276329239.&      276329188.&      276329181.&      276328153.&      276328147.\\
&&&          92.166$^\circ$&          92.039$^\circ$&          88.282$^\circ$&          88.218$^\circ$&          87.924$^\circ$&          87.882$^\circ$&          82.169$^\circ$&          82.139$^\circ$\\ \hline
RS Cnc&087&e&      276334231.&      276334252.&      276334889.&      276334900.&      276334951.&      276334958.&      276335986.&      276335992.\\
&&&         119.093$^\circ$&         119.220$^\circ$&         122.976$^\circ$&         123.040$^\circ$&         123.334$^\circ$&         123.376$^\circ$&         129.088$^\circ$&         129.118$^\circ$\\ \hline
$\gamma$ Cru&089&i&      277408171.&      277408164.&      277407931.&      277407926.&      277407905.&      277407902.&      277407406.&      277407403.\\
&&&         177.322$^\circ$&         177.330$^\circ$&         177.611$^\circ$&         177.617$^\circ$&         177.641$^\circ$&         177.645$^\circ$&         178.207$^\circ$&         178.211$^\circ$\\ \hline
$\gamma$ Cru&093&i&      280044544.&      280044536.&      280044270.&      280044265.&      280044241.&      280044238.&      280043673.&      280043670.\\
&&&         207.566$^\circ$&         207.558$^\circ$&         207.301$^\circ$&         207.296$^\circ$&         207.273$^\circ$&         207.270$^\circ$&         206.755$^\circ$&         206.752$^\circ$\\ \hline
$\gamma$ Cru&094&i&&&      280680556.&      280680551.&      280680529.&      280680526.&&\\
&&&&&         191.749$^\circ$&         191.749$^\circ$&         191.751$^\circ$&         191.751$^\circ$&&\\ \hline
$\gamma$ Cru&096&i&      282013705.&&      282013475.&      282013471.&      282013451.&      282013448.&      282012973.&      282012970.\\
&&&         185.519$^\circ$&&         185.647$^\circ$&         185.649$^\circ$&         185.660$^\circ$&         185.662$^\circ$&         185.911$^\circ$&         185.912$^\circ$\\ \hline
$\gamma$ Cru&100&i&      285033365.&      285033358.&      285033091.&      285033085.&      285033061.&      285033058.&      285032494.&      285032491.\\
&&&         222.666$^\circ$&         222.648$^\circ$&         222.045$^\circ$&         222.034$^\circ$&         221.980$^\circ$&         221.973$^\circ$&         220.775$^\circ$&         220.768$^\circ$\\ \hline
$\gamma$ Cru&101&i&      285860519.&      285860511.&      285860244.&      285860239.&      285860215.&      285860211.&      285859648.&      285859644.\\
&&&         222.671$^\circ$&         222.653$^\circ$&         222.050$^\circ$&         222.038$^\circ$&         221.985$^\circ$&         221.977$^\circ$&         220.779$^\circ$&         220.772$^\circ$\\ \hline
$\gamma$ Cru&102&i&      286685692.&      286685684.&      286685418.&      286685413.&      286685389.&      286685386.&&\\
&&&         222.343$^\circ$&         222.326$^\circ$&         221.730$^\circ$&         221.718$^\circ$&         221.665$^\circ$&         221.658$^\circ$&&\\ \hline
$\beta$ Peg&104&i&&      288914069.&      288913926.&      288913923.&      288913911.&      288913909.&      288913607.&      288913605.\\
&&&&         344.206$^\circ$&         344.808$^\circ$&         344.819$^\circ$&         344.872$^\circ$&         344.880$^\circ$&         346.075$^\circ$&         346.082$^\circ$\\ \hline
R Cas&106&i&      291037657.&      291037640.&      291037086.&      291037076.&      291037029.&      291037023.&      291036027.&      291036022.\\
&&&          79.815$^\circ$&          79.727$^\circ$&          76.965$^\circ$&          76.916$^\circ$&          76.687$^\circ$&          76.655$^\circ$&          71.974$^\circ$&          71.949$^\circ$\\ \hline
$\alpha$ Sco&115&i&&&      302021191.&      302021182.&      302021136.&      302021130.&      302020071.&      302020065.\\
&&&&&         160.460$^\circ$&         160.473$^\circ$&         160.534$^\circ$&         160.543$^\circ$&         161.910$^\circ$&         161.918$^\circ$\\ \hline
$\beta$ Peg&170&e&      397974243.&      397974253.&&&      397974644.&      397974648.&&\\
&&&          79.751$^\circ$&          79.765$^\circ$&&&          80.298$^\circ$&          80.304$^\circ$&&\\ \hline
$\beta$ Peg&172&i&      401619627.&&      401619261.&      401619254.&&&      401618465.&      401618460.\\
&&&         310.964$^\circ$&&         310.439$^\circ$&         310.429$^\circ$&&&         309.362$^\circ$&         309.356$^\circ$\\ \hline
W Hya&179&i&      411902470.&      411902464.&      411902265.&      411902261.&&&&\\
&&&         144.025$^\circ$&         143.998$^\circ$&         143.082$^\circ$&         143.065$^\circ$&&&&\\ \hline
W Hya&180&i&      413051882.&      413051876.&      413051675.&      413051671.&      413051653.&      413051651.&      413051233.&      413051230.\\
&&&         144.512$^\circ$&         144.484$^\circ$&         143.551$^\circ$&         143.533$^\circ$&         143.451$^\circ$&         143.439$^\circ$&         141.610$^\circ$&         141.599$^\circ$\\ \hline
W Hya&181&i&      414201120.&      414201114.&      414200913.&      414200909.&      414200891.&      414200889.&      414200471.&      414200468.\\
&&&         144.580$^\circ$&         144.552$^\circ$&         143.616$^\circ$&         143.598$^\circ$&         143.516$^\circ$&         143.504$^\circ$&         141.668$^\circ$&         141.658$^\circ$\\ \hline
R Cas&185&i&      418066525.&&&&&&&\\
&&&         328.466$^\circ$&&&&&&&\\ \hline
$\mu$ Cep&185&e&      418017173.&      418017190.&      418017754.&      418017765.&      418017815.&      418017822.&      418018966.&      418018973.\\
&&&          48.225$^\circ$&          48.267$^\circ$&          49.666$^\circ$&          49.692$^\circ$&          49.814$^\circ$&          49.831$^\circ$&          52.503$^\circ$&          52.518$^\circ$\\ \hline
$\gamma$ Cru&187&i&      419919080.&      419919073.&      419918807.&      419918802.&      419918778.&      419918775.&      419918229.&      419918226.\\
&&&         148.447$^\circ$&         148.410$^\circ$&         147.187$^\circ$&         147.164$^\circ$&         147.057$^\circ$&         147.042$^\circ$&         144.678$^\circ$&         144.664$^\circ$\\ \hline
$\gamma$ Cru&187&e&      419931192.&      419931200.&      419931465.&      419931470.&      419931494.&      419931497.&      419932041.&      419932044.\\
&&&         228.830$^\circ$&         228.867$^\circ$&         230.087$^\circ$&         230.110$^\circ$&         230.217$^\circ$&         230.232$^\circ$&         232.590$^\circ$&         232.604$^\circ$\\ \hline
$\mu$ Cep&191&i&      423055904.&      423055895.&      423055573.&      423055567.&      423055538.&      423055533.&      423054848.&      423054844.\\
&&&         289.807$^\circ$&         289.802$^\circ$&         289.625$^\circ$&         289.621$^\circ$&         289.605$^\circ$&         289.603$^\circ$&         289.248$^\circ$&         289.246$^\circ$\\ \hline
$\mu$ Cep&193&i&      425122618.&      425122608.&      425122286.&      425122280.&      425122251.&      425122247.&      425121561.&      425121557.\\
&&&         289.993$^\circ$&         289.987$^\circ$&         289.806$^\circ$&         289.802$^\circ$&         289.786$^\circ$&         289.784$^\circ$&         289.420$^\circ$&         289.417$^\circ$\\ \hline
R Cas&194&e&&&      426260787.&      426260792.&&&&\\
&&&&&          85.118$^\circ$&          85.124$^\circ$&&&&\\ \hline
\end{tabular}}
\end{table}

\begin{table}
\caption{Time, longitude and phase differences used to determine pattern speeds (full table included in on-line supplement)}
\label{phasetab}
\centerline{\begin{tabular}{|c|c|c|c|c|c|} \hline 
Wave  & Occultation Pair       &    $\delta t$ (days) &   $\delta\lambda$  (deg.) & $\delta\phi$ (deg.) &   $\sigma_\phi$ (deg.) \\ \hline
W82.00 & RSCnc085e-RSCnc085i  &        0.03051 &       16.9   &    241.2  &       2.7 \\
W82.06 & RSCnc085e-RSCnc085i   &       0.03137 &       17.4   &    242.2   &      6.0 \\
W82.21 & RSCnc085e-RSCnc085i&          0.03381 &       18.7   &    242.4   &      7.2 \\
W83.63 & RSCnc087e-RSCnc087i&          0.03478 &       18.9   &    48.5   &       7.1\\
W84.64 & RSCnc087e-RSCnc087i&          0.04799 &       25.9   &    223.8   &      5.0 \\
\hline
\end{tabular}}
\end{table}

\begin{table}
\caption{Summary of wavelet analyses}
\label{sumtab}
\resizebox{6in}{!}{\begin{tabular}{|c|c|c|c|c|c|c|c|c|c|c|}\hline
Wave & Resonant & Region & Wavelengths & $N(\delta \phi)^c$ & m & Pattern & Rotation  & $\sigma_0$ & $\tau_n^a$ & $\tau_n/\sigma_0$ \\
& Location$^a$ & Considered$^b$ & Considered$^b$ & & & Speed & Period & & & \\ \hline
W80.98 & 80,998 km & 80970-80995 km & 0.1-5 km & 193 & -4 & 1660.4$^\circ$/day & 312.2 min & 5.85 g/cm$^2$ & 0.13 & 0.022 cm$^2$/g \\
W82.00 & 82,010 km & 81992-82012 km & 0.1-5 km & 255 & -3 & 1736.6$^\circ$/day & 298.5 min & 5.68 g/cm$^2$ & 0.14 & 0.025 cm$^2$/g \\ 
W82.06 & 82,061 km & 82040-82065 km & 0.1-5 km & 286 & -3 & 1735.0$^\circ$/day & 298.8 min & 10.16 g/cm$^2$ & 0.28 & 0.028 cm$^2$/g \\
W82.21 & 82,209 km & 82190-82215 km & 0.1-5 km & 247 & -3 & 1730.3$^\circ$/day & 299.6 min & 6.92 g/cm$^2$ & 0.13 & 0.020 cm$^2$/g \\
W83.63 & 83,633 km & 83625-83635 km & 0.1-5 km & 63 & -10 & 1394.1$^\circ$/day & 371.9 min & 4.95 g/cm$^2$ &  0.10 & 0.020 cm$^2$/g \\ 
W84.64 & 84,644 km & 84625-84650 km & 0.1-5 km & 311 & -2 & 1860.8$^\circ$/day & 278.6 min & 4.05 g/cm$^2$ & 0.11 & 0.027 cm$^2$/g \\
W84.82 & 84,814 km & 84810-84830 km & 1.0-5 km & 218 & +3 & 833.5$^\circ$/day &  622.0 min & 3.94  g/cm$^2$ & 0.44 & 0.11 cm$^2$/g \\
W84.86 & 84,857 km & 84850-84880 km & 1.0-5 km & 146 & +3 & 833.0$^\circ$/day & 622.3 min & 2.24  g/cm$^2$ &  0.42 & 0.19 cm$^2$/g \\
W85.67 & 85,677 km & 85675-85690 km & 1.0-5 km & 300$^d$ & -1 & 2430.5$^\circ$/day & 213.3 min & --- & 0.29 & --- \\
W86.40 & 86,400 km & 86400-86420 km & 1.0-5 km & 179 & +3 & 810.4$^\circ$/day & 639.7 min & 4.70  g/cm$^2$ & 0.47  & 0.10 cm$^2$/g \\
W86.58 & 86,576 km & 86575-86585 km & 1.0-5 km & 121 & +3 & 807.9$^\circ$/day & 641.7 min & 1.18  g/cm$^2$ & 0.38  & 0.31 cm$^2$/g \\
W86.59 & 86,590 km & 86595-86605 km & 1.0-5 km$^e$ & 243 & +3 & 807.7$^\circ$/day & 641.8 min &  --- & ---  & --- \\
W87.19 & 87,189 km & 87175-87205 km & 0.1-5 km & 164 & -2 & 1779.5$^\circ$/day & 291.3 min & 1.41g/cm$^2$ & 0.15 & 0.11 cm$^2$/g \\
\hline
\end{tabular}}
$^a$ From Table 7 of \citet{Baillie11}

$^b$ Radial range and wavelength range considered in wavelet analysis.

$^c$ Number of $\delta \phi$ values used in fit.

$^d$ Includes observations with $\delta t$ as large at 1000 days.

$^e$ Data high-pass filtered prior to fitting.

\end{table}

\section{W83.63: An $m=-10$ wave}
\label{W83.63}

\begin{figure}
\centerline{\resizebox{3in}{!}{\includegraphics{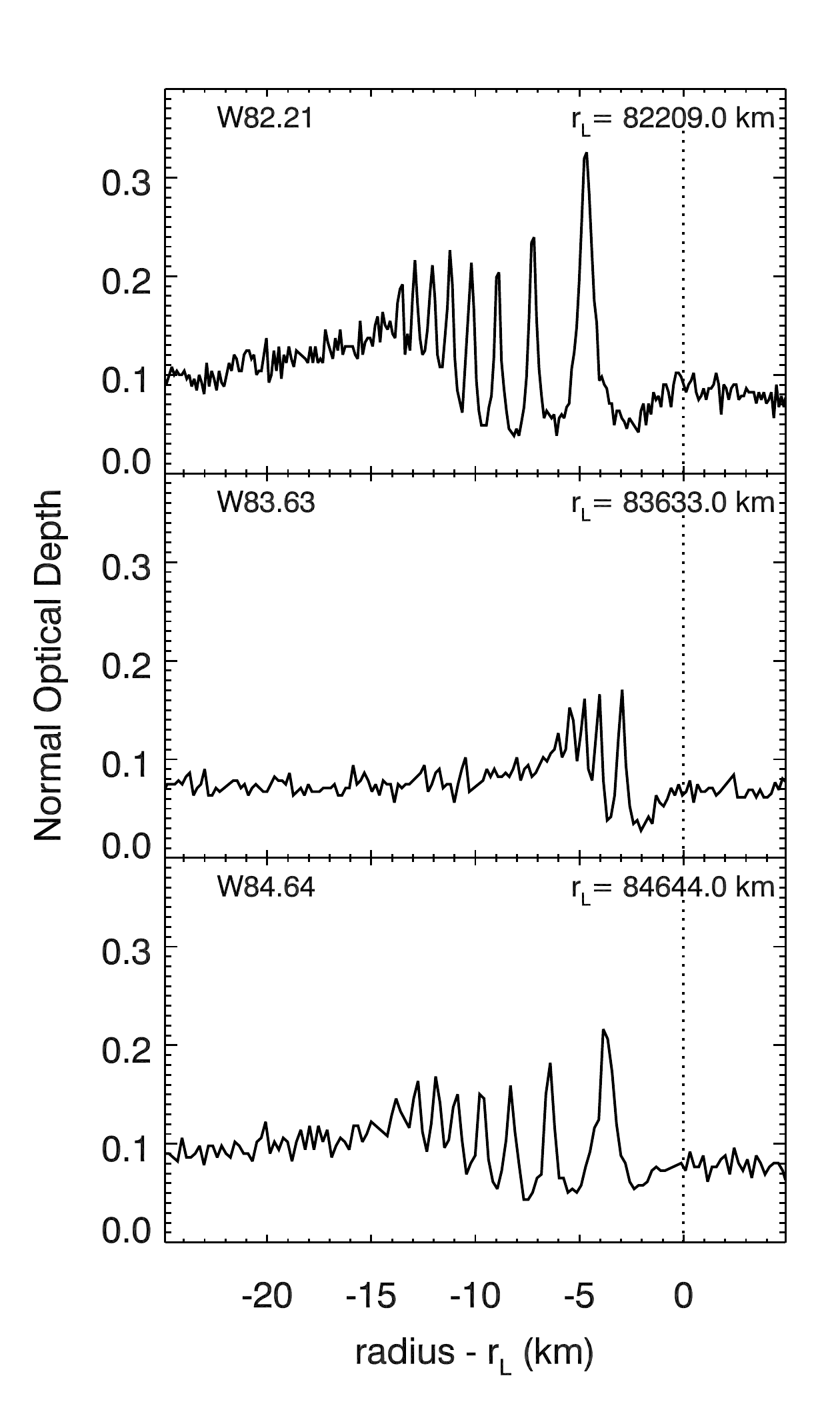}}}
\caption{Sample Profile of W83.63, compared with two other inward-propagating waves in its vicinity. These
profiles are from the egress portion of the Rev 85 RSCnc occultation (B=29.96$^\circ$) which sampled all three
waves relatively well. \citet{HN13} showed that W82.21 is likely an $m=-3$ wave, while W84.64 is probably an $m=-2$ wave. Note that W83.63 has a noticeably shorter wavelength than either of the other waves, but its opacity variations are close to saturation, like the other two waves.}
\label{W83.63prof}
\end{figure}

\begin{figure}
\resizebox{6in}{!}{\includegraphics{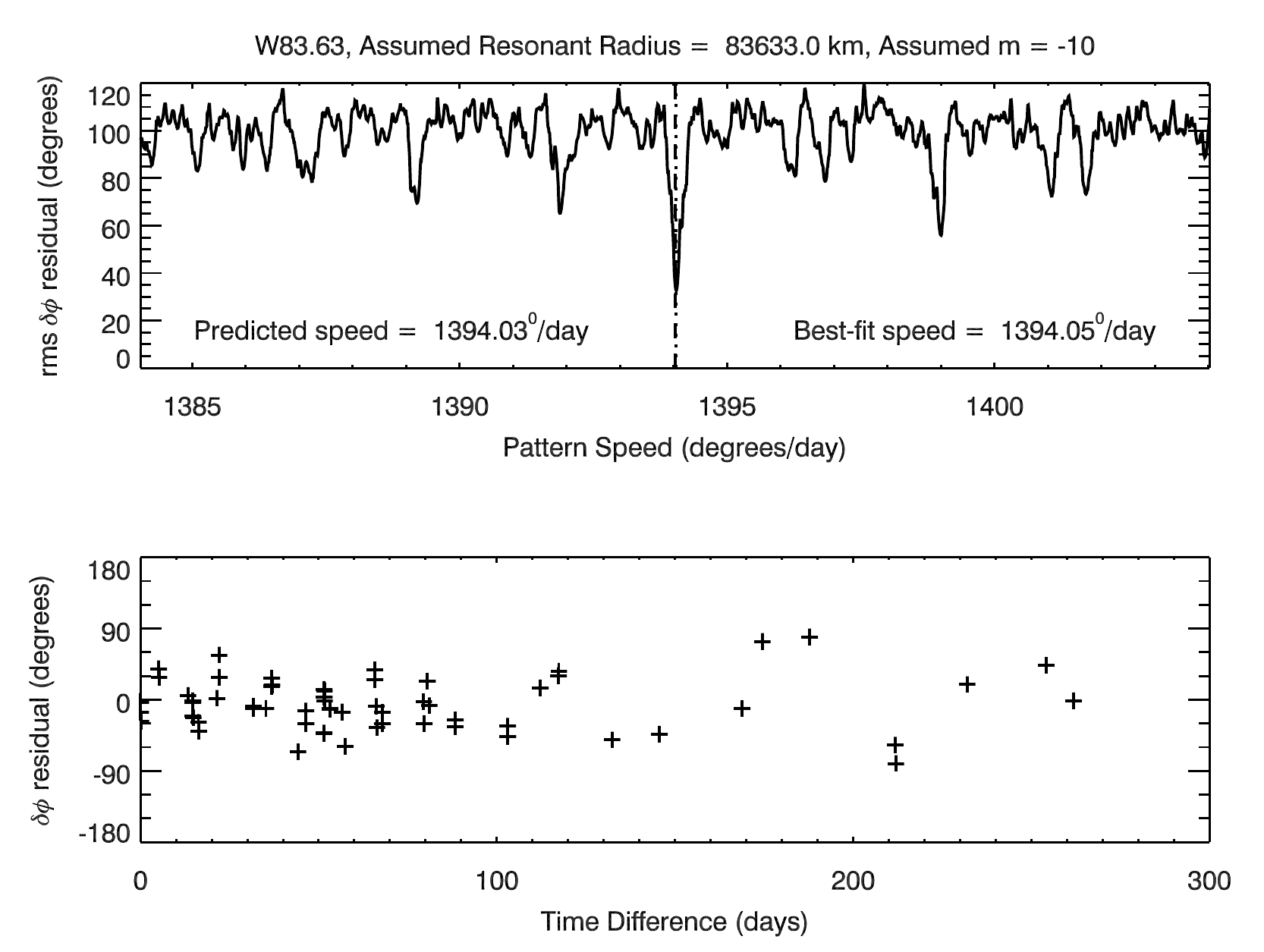}}
\caption{Results of a wavelet analysis of wave W83.63 which considered the radial range of 83,625-83,635 km and a pattern wavelength range of 0.1 to 5 km. The top panel shows the $rms$ phase difference residuals as a function of  pattern speed, assuming the wave is an $m=-10$ pattern. The dashed line marks the expected
pattern speed for such a structure with the appropriate resonant radius, while the dotted line indicates the best-fit pattern speed. There is a clear minimum in the residuals very close to the predicted location (other minima are likely aliasing within the sparse data set). The bottom panel shows the phase difference residuals (observed-expected) for this best-fit solution as a function of time difference between the observations. Note the scatter in the residuals is around 45$^\circ$.}
\label{W83.63res}
\end{figure}

The first wave we will consider here is the one designated W83.63 in \citet{Colwell09} (it is also called $h$ in \citet{Rosen91} and wave 18 in \citet{Baillie11}). An example profile of this wave is shown in Figure~\ref{W83.63prof}, where it is compared with two nearby waves examined by \citet{HN13}. Note this wave is found in a low optical-depth region between three waves (W82.00, W82.06 and W82.21) that \citet{HN13} identified as $m=-3$ and another wave (W84.64) that appears to have $m=-2$ (see Figure~\ref{overview}). As shown in Figure~\ref{W83.63prof}, W83.63 has a much shorter wavelength than either the $m=-3$ or the $m=-2$ waves. W83.63's  shorter wavelength means that some of the profiles that could provide useful data on the other waves do not have sufficient resolution to yield sensible phase information for this wave. Furthermore, its shorter wavelength will also make the phase measurements more sensitive to slight geometry errors. Hence we elected not to look at this wave in our initial study. However, subsequent investigations revealed nineteen VIMS occultations with sufficient resolution to yield useful phase data for this wave (see Table~\ref{obstab}), and trajectory refinements provided by R. French were sufficiently precise to yield a sensible solution for $m$ and $\Omega_p$. Note that when we computed phase differences between the wave profiles, we considered wavelengths between 0.1 and 5 km.

Since the wave appears to propagate inwards, like the waves described in \citet{HN13}, we expected that this wave would have a negative value for $m$. Indeed, when we examined a range of different $m$-values, we found the smallest dispersion in the phase residuals occurred with $m=-10$ and a pattern speed close to the expected value for an $m=-10$ wave with a resonant radius $r_L=83633$ km (the estimated resonant radius from Baillie {\it et al.} 2011). A profile of the $rms$ residuals as a function of pattern speed (assuming $m=-10$) is provided in Figure~\ref{W83.63res}, which shows that the best-fit solution corresponds to $\Omega_p=1394.05^\circ$/day. This solution has an $rms$ residual of around 45$^\circ$, roughly two times larger than the scatter around the best-fit solutions for the better-resolved waves examined in \citet{HN13}. Nevertheless, the residuals (also shown in Figure~\ref{W83.63res}) exhibit a reasonably tight distribution around zero with no obvious trends out to 300 days. 

Additional, weaker minima can be observed in the $rms$ residuals in the upper panel of Figure~\ref{W83.63res}. Also, if we assume $m=-9$, we found a single minimum where the $rms$ reaches about $60^\circ$ for a value of  $\Omega_p\sim$0.2$^\circ$/day away from the expected pattern speed.  All of these other solutions are substantially worse than the  $m=-10$, $\Omega_p=1394.05^\circ$/day solution, and so we do not regard them as valid alternatives. Instead, they are likely due to aliasing among the relatively sparse observations. We have also confirmed that $m=-10$, $\Omega_p=1394.05^\circ$/day remains the best solution even if we add other observations with somewhat poorer resolution and/or data gaps. Thus we are fairly confident that W83.63 is indeed an $m=-10$ wave.

We may also note that a value of $m=-10$ is plausible based on a consideration of the rings' local surface mass density. The wavelength of a given density wave depends on both the $m$-value and the (unperturbed) ring surface mass density $\sigma_0$, and \citet{Baillie11} provide scaled estimates of  the surface mass density $\sigma_0=0.45|m-1|$ g/cm$^2$ and the opacity $\tau_n/\sigma_0=0.22/|m-1|$ cm$^2$/g based on their observations of this wave.  If we insert $m=-10$ into these expressions, then we obtain a surface mass density of 4.95 g/cm$^2$ and an opacity of 0.020 cm$^2$/g for this wave. These numbers are comparable to the values obtained from the nearby $m=-2, -3$ and $m=-4$ waves  (see Table~\ref{sumtab} and Figure~\ref{sigma} below).

The derived value for $m$ is also reasonably consistent with the theoretical predictions by \citet{MarleyPorco93}. While those authors only predicted resonance locations for  patterns with $|m|$ between 2 and 8,  one can extrapolate from these data to estimate that the resonance with $|m|=10$  should indeed fall somewhere in the vicinity of W83.63. Combined with the reasonable surface mass density estimate, and the close match between the observed and expected pattern speeds, this makes the identification of W83.63 as an $m=-10$ wave reasonably secure.

The existence of this clear $m=-10$ wave raises some interesting questions about the relative amplitudes of the waves generated by normal modes inside the planet. According to \citet{MarleyPorco93}, each fundamental sectoral (i.e. $\ell=m$) normal mode within the planet gives rise to a wave in the rings with the same azimuthal wavenumber $|m|$, and the amplitude of the wave should depend upon the magnitude of the corresponding planetary normal mode. \citet{HN13} found $m=-2$, -3 and -4 waves with substantial amplitudes in the middle C ring, which correspond to  $m=2, 3$ and 4 fundamental normal modes within the planet. Now we have another wave with a large amplitude ($\delta \tau/\tau \simeq 1$) that appears to be generated by the  $m=10$ fundamental sectoral normal mode. However, resonances with the $m=5$ through $m=9$ normal modes should also fall between 80,000  and 84,000 km, and there are no other waves with amplitudes comparable to W83.63 in this region. This is especially surprising because this entire region is largely featureless outside of the relevant waves (see Figure~\ref{overview}), so it is unlikely that these waves would be lost or obscured by other structures. While further analysis could potentially reveal weak waves within this region, the available data  suggest that the $m$=5-9 planetary normal modes have significantly lower amplitudes than the $m=10$ mode. This is inconsistent with the expected spectrum of normal modes computed by Marley \& Porco (1993, Fig. 6) assuming simple mode-energy partition schemes \citep{GK88, GK90}, and implies that the excitation spectrum of the planetary normal modes has a non-trivial shape.

\section{W81.02, another wave generated by a normal mode oscillation?}
\label{W81.02}

\begin{figure}
\centerline{\resizebox{3in}{!}{\includegraphics{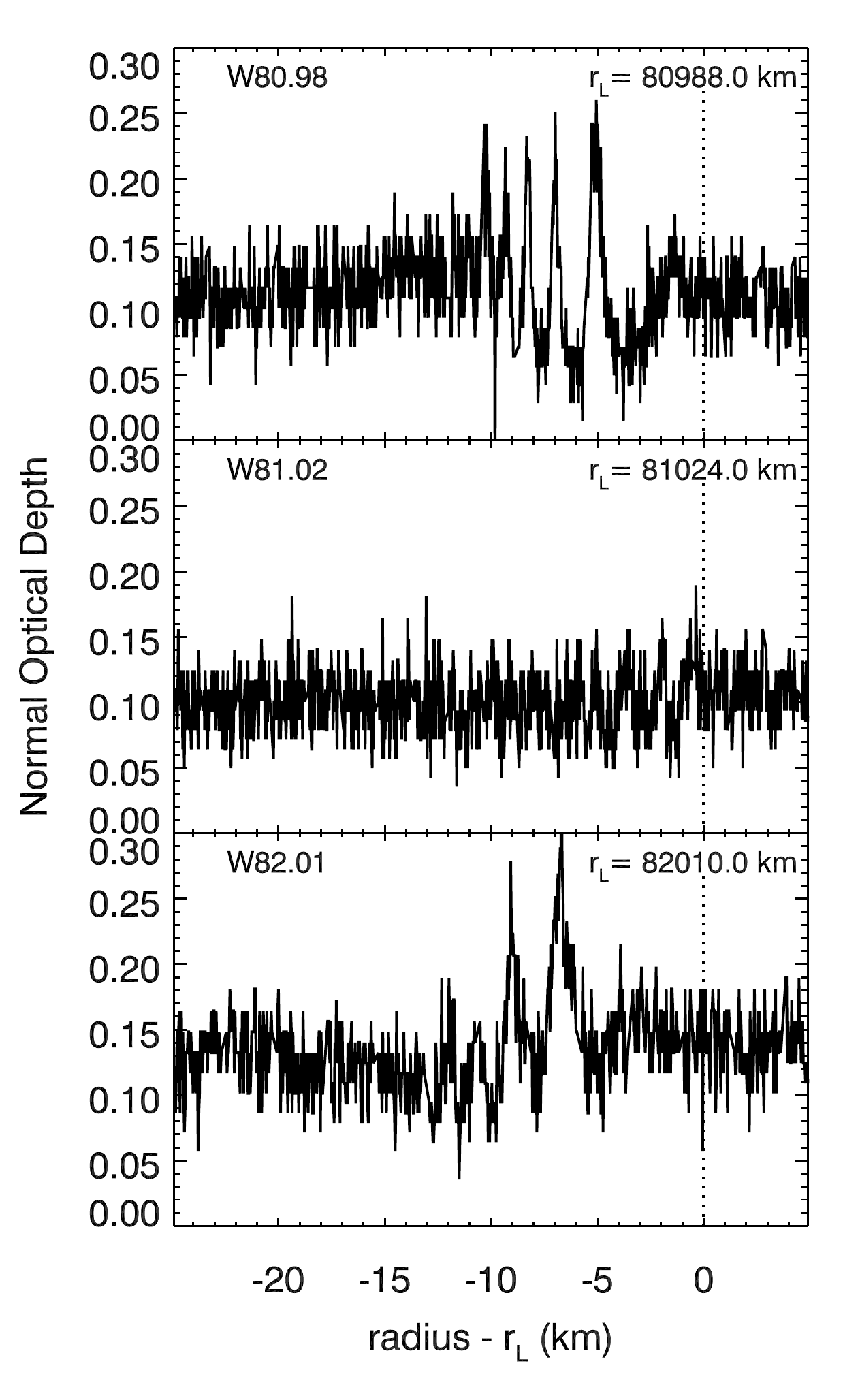}}}
\caption{Sample Profile of W81.02, compared with two other inward-propagating waves in its vicinity. These
profiles are from the Rev 106 RCas occultation (B=56.04$^\circ$) which sampled all three
waves relatively well. \citet{HN13} showed that W80.98 is likely an $m=-4$ wave, while $W82.01$ is probably an $m=-3$ wave. Note that \citet{Baillie11} identified W81.02 as an outward-propagating wave with a resonance location around 81,018 km. To facilitate comparisons, we here plot the wave as if it were an inward-propagating wave initiated
at 81,024 km. This wave is clearly much weaker than the other two, and the wavelength trend with radius is ambiguous.}
\label{W81.02prof}
\end{figure}

\begin{figure}
\resizebox{6in}{!}{\includegraphics{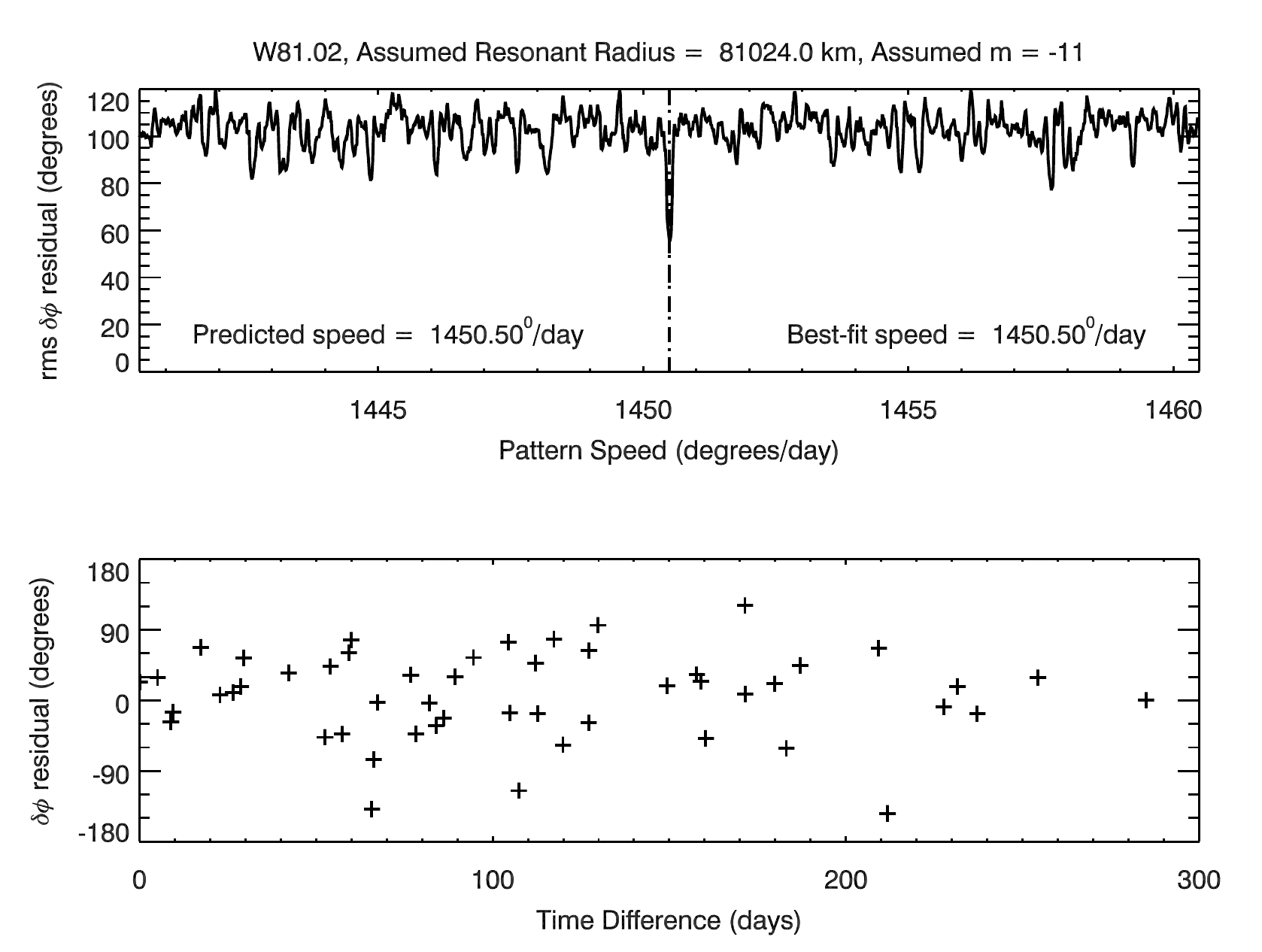}}
\caption{Results of a wavelet analysis of wave W81.02 which considered the radial range of 81,010-81,030 km and a pattern wavelength range of 0.5 to 2 km. The top panel shows the $rms$ phase difference residuals as a function of  pattern speed, assuming the wave is an $m=-11$ pattern. The dashed line marks the expected
pattern speed for such a structure with the appropriate resonant radius. There is a minimum in the residuals very close to the predicted location. The bottom panel shows the phase difference residuals (observed-expected) for this best-fit solution as a function of time difference between the observations.}
\label{W81.02res}
\end{figure}

\begin{figure}
\resizebox{6in}{!}{\includegraphics{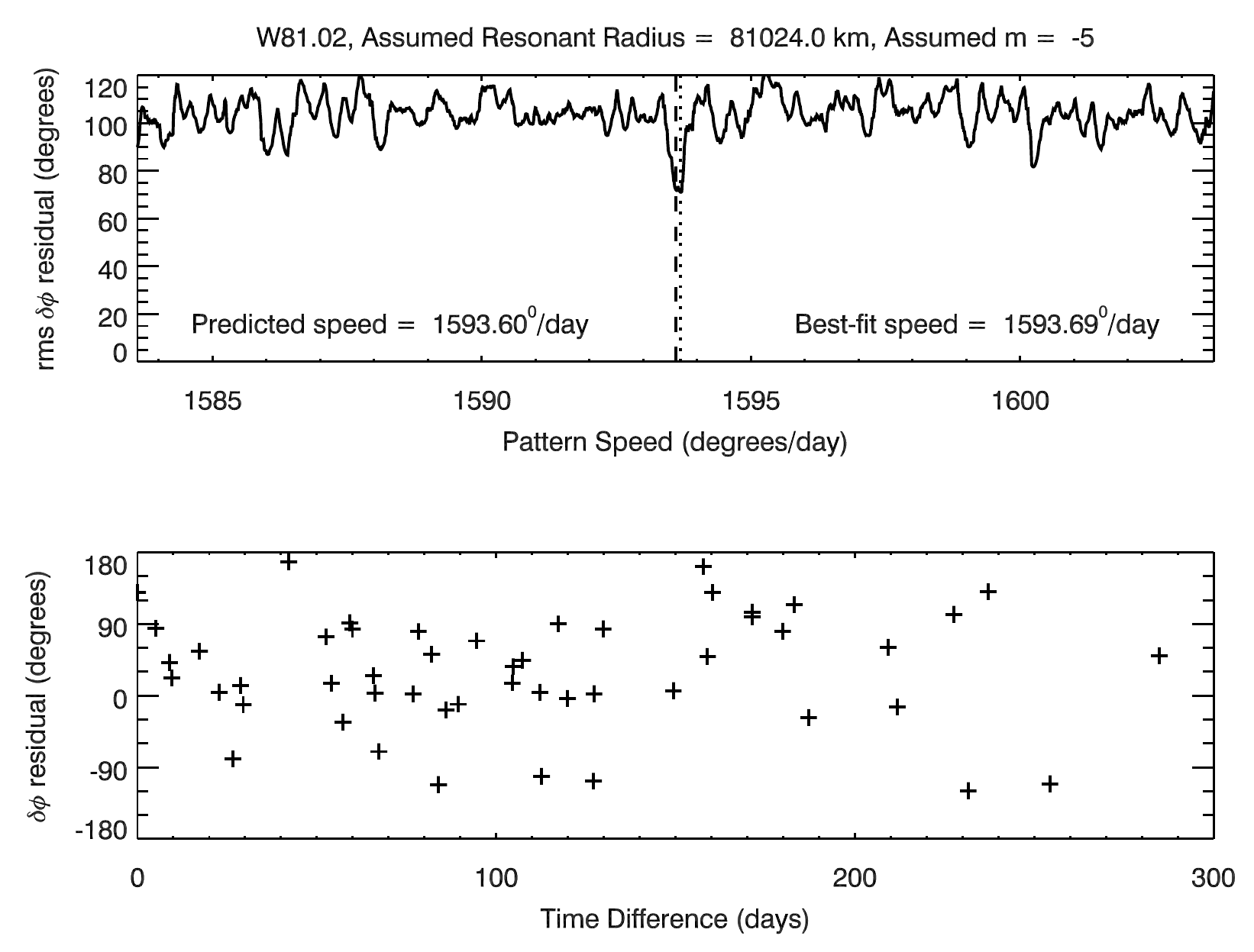}}
\caption{Results of a wavelet analysis of wave W81.02 which considered the radial range of 81,010-81,030 km and a pattern wavelength range of 0.5 to 2 km. The top panel shows the $rms$ phase difference residuals as a function of  pattern speed, assuming the wave is an $m=-5$ pattern. The dashed line marks the expected
pattern speed for such a structure with the appropriate resonant radius. There is a weak minimum in the residuals close to the predicted location. The bottom panel shows the phase difference residuals (observed-expected) for this alternate solution as a function of time difference between the observations.}
\label{W81.02res2}
\end{figure}

\citet{Baillie11} identified a rather weak feature in the middle C ring that could potentially be generated by another planetary normal mode. They designated this feature with the number 14, but to be consistent with our existing notation and \citet{Colwell09} we will call this wave W81.02 here. Figure~\ref{W81.02prof} shows a sample profile of this wave, compared with the profiles of two normal-mode generated waves previously analyzed by \citet{HN13}. W81.02 is clearly much weaker than the other waves we have analyzed. Indeed, it is difficult to discern whether this wave propagates inwards or outwards. Based on co-added data from multiple occultations, \citet{Baillie11} infer that this is an outward-propagating wave, but given the feature's low signal-to-noise, this finding is not as secure as it is for other waves in this region.

We considered both positive and negative values of $m$ in our analysis of this wave. We found no clear minimum at the appropriate pattern speed for any positive value of $m$, but at least two negative values of $m$ yielded a minimum near the expected pattern speeds. If we considered a radial range of 81,010-81,030 km and a wavelength range of 0.5-2.0 km, the deepest minimum was found with $m=-11$ (see Figure~\ref{W81.02res}). However, a weak minimum also exists for $m=-5$ (see Figure~\ref{W81.02res2}). In principle, the relative depths of the two minima could be used to estimate the relative likelihood that either one of these solutions is correct. However, in practice the depths of the minima are sensitive to the ranges of wavelengths and radii considered in the analysis, which makes the relative statistical significance of these solutions difficult to quantify. Thus we will simply note that the dispersion in the phase residuals around the $m=-5$ solution is larger than it is for the $m=-11$ solution, and  that the $m=-5$ solution is less robust against small changes in the ranges of wavelengths and radii considered in the analysis. Both of these considerations would lead us to favor the $m=-11$ solution, but  we cannot definitively rule out the $m=-5$ solution at this point.

While our analysis was able to provide two potential solutions, at present we regard these identifications as extremely tentative. Not only is the signal-to-noise of this feature extremely low and the dispersion in the phase residuals high, but we also cannot use the inferred surface mass density and resonance location to confirm the identity  of this wave.  A negative value for $m$ implies that the feature is an inward-propagating wave, contradicting  the \citet{Baillie11} analysis. While \citet{Baillie11} could have misidentified the wave as outward-propagating on account of its low signal-to-noise and limited radial extent, the lack of consistency between our findings means that we cannot use their estimates of the ring's opacity or surface mass density to confirm our estimate of $m$. 

If we assume that W81.02 is generated by a fundamental sectoral normal mode inside Saturn, then the location of this feature could in principle help confirm its identification. According to \citet{MarleyPorco93}, the locations of the sectoral normal modes follow a non-monotonic but well-defined trend that arises because both the predicted pattern speeds for the sectoral normal modes and the resonant pattern speeds for outer Lindblad resonances at a given radius decrease with increasing $|m|$. For $m$ between 2 and 5, it turns out that the resonant locations shift inwards with increasing $m$ from about 86,000 km to 81,000 km. For higher values of $m$ the resonant locations steadily move outwards with increasing $m$. The locations and pattern speeds of the other waves in this region follow this predicted  trend (see Figure~\ref{reslocs}). W81.02 falls just exterior to W80.98, which \citet{HN13} identified as an $m=-4$ wave, and interior to the $m=-10$ wave W83.63. Hence W81.02 falls close to the predicted resonance location for the $m=5$ fundamental sectoral normal mode, and well interior to the predicted resonance location for the $m=11$ normal mode. Such considerations would favor the $m=-5$ solution for this wave. However,  this argument presupposes that this pattern should occur at one of the predicted resonant locations for fundamental sectoral normal modes. Even if this wave is generated by such an oscillation, differential rotation could displace the predicted resonant locations inwards \citep{Marley91}.  Barring more detailed models of the planet's internal oscillations, we cannot yet rule out the possibility that a $m=11$ pattern could have a resonance near W81.02. 

Since the dispersion in phase measurements favors the $m=-11$ solution, while its location could support the $m=-5$ option, it would be premature to categorize W81.02 as either pattern. Hence we do not list this feature in Table~\ref{sumtab}. However, this wave does merit further investigation with larger data sets and more refined analytical methods, since it could  represent another resonance with planetary normal-mode oscillations.

\section{Multiple $m=+3$ waves in the outer C ring}
\label{m3waves}

\begin{figure}
\centerline{\resizebox{6in}{!}{\includegraphics{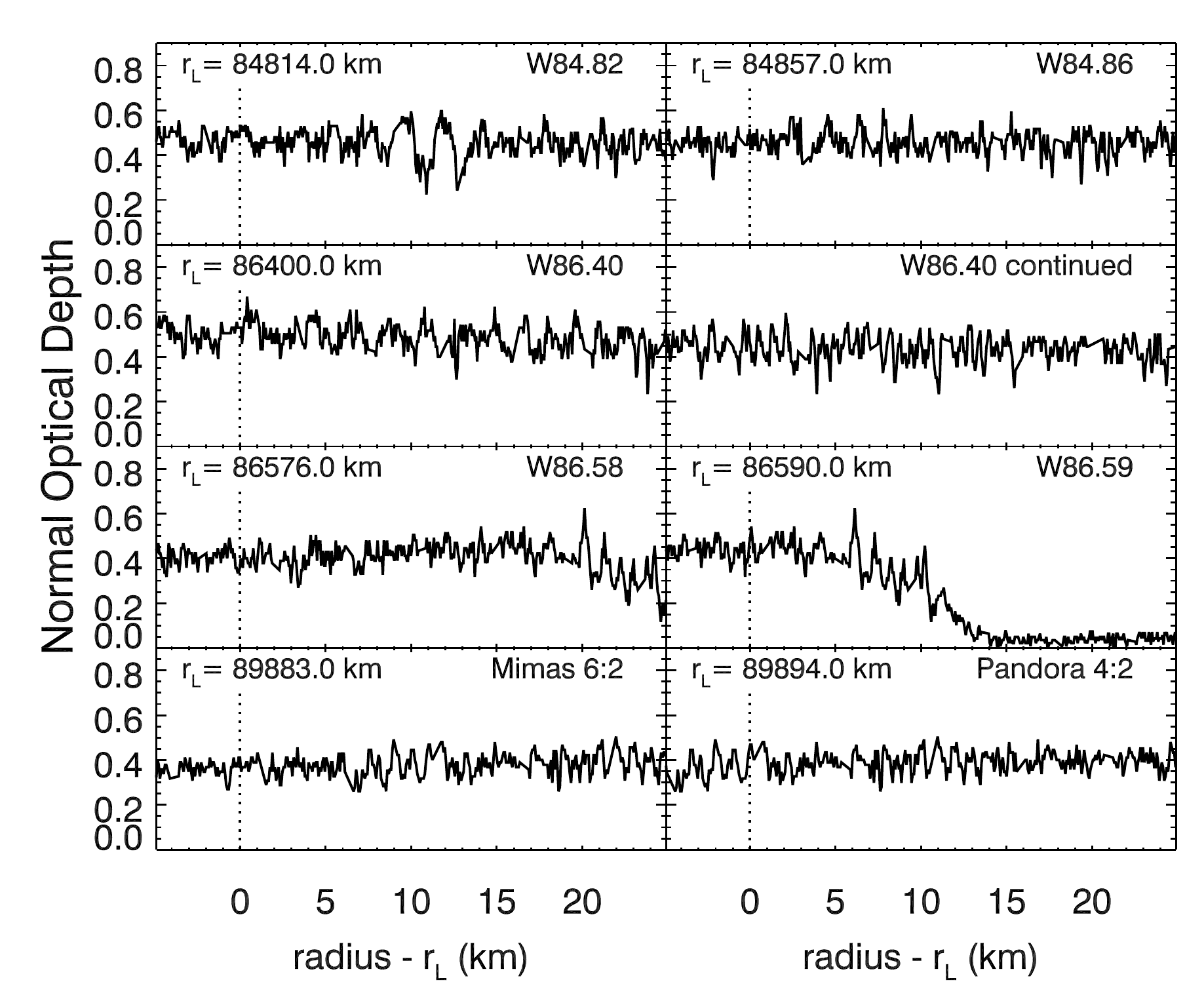}}}
\caption{Sample profiles of seven C-ring waves from the Rev 106 RCas occultation (B=56.04$^\circ$). Vertical dotted lines indicate the resonant radii, as fitted by \citet{Baillie11}. Note that these waves are all found in plateaux, and have much lower signal-to-noise than the waves illustrated in Figure~\ref{W83.63prof}. Still, quasi-periodic patterns  with wavelengths around 1 km  can be seen extending outwards of all the marked locations.}
\label{m3prof}
\end{figure}

Besides W81.02, seven of the weak waves discovered by \citet{Baillie11} were detectable in individual VIMS occultations profiles. These were designated features 20, 21, 28, 29, 31, 36 and 37 by \citet{Baillie11}, who suggested that the outermost two of these features are generated by the  Mimas 6:2 and Pandora 4:2 inner Lindblad resonances at  89,883 km and 89,894 km, respectively. They also point out that  feature 20 is  only a few kilometers from the Pan 4:2 inner Lindblad resonance, but the distance between the wave and the resonance, along with lack of an obvious wave at the much stronger 2:1 resonance, strongly suggests that this is just a coincidence. Hence \citet{Baillie11} conclude that the inner five waves have no known resonance that could explain them. To be consistent with the nomenclature of other unknown features \citep{Colwell09}, we will here designate these features as waves W84.82, W84.86, W86.40, W86.58 and W86.59  (the nominal resonant locations of the inner four features are 84,814 km, 84,857 km, 86,400 km and 86,576 km, respectively according to \citet{Baillie11}, while the last feature appears to have a resonant radius around 86,590 km).

Figure~\ref{m3prof} shows sample profiles of these seven waves. While some periodic structures are visible at these locations, the signal-to-noise of these features is much lower than it was for the waves considered by  \citet{HN13}. This is not only because the amplitudes of the waves themselves are smaller, but also because they occupy  sharp-edged regions of enhanced optical depth known as plateaux (see Figure~\ref{overview}). These regions show enhanced fine-scale stochastic optical depth variations compared with the lower optical-depth parts of the C ring, and this fine-scale structure further obscures the waves. By combining  wavelet data from multiple occultations,  \citet{Baillie11}  found that, unlike the waves analyzed by \citet{HN13}, all of these  waves appear to propagate {\it outwards}. Assuming these features are density waves,\footnote{While vertical bending waves with pattern speeds faster than the local mean motion would also propagate outwards \citep{Shu84}, the observed properties of the waves discussed here are not consistent with bending waves. The apparent optical depth contrast of vertical structures like bending waves depends on the elevation angle of the line-of-sight to the star through the ringplane, but the appearance of these waves exhibits no obvious trends with elevation angle. Also, the pattern speeds of density and bending waves are quite different, and the best-fit pattern speeds of these features are consistent with density waves.}  this implies that their pattern speeds are slower than the local mean motion. Hence they are likely driven by inner Lindblad resonances and should all have a positive $m$-values.

Table~\ref{obstab3} provides a list of the occultations with sufficient resolution and coverage to yield decent phase information on these waves. While the numbers of occultations available are comparable to those used in our earlier analyses of other waves, it was not immediately clear if our algorithms could provide robust estimates of pattern speeds and symmetry properties for such low amplitude features. Fortunately, we can actually test our algorithms using the two features \citet{Baillie11} attribute to the 6:2 Mimas and 4:2 Pandora Lindblad resonances, both of which are located in Plateau P10 between 89,800 km and 89,900 km. If these identifications are correct, then we would expect both these structures to be  $m=+3$ patterns with the appropriate pattern speeds.  Note that  these two satellite-induced waves are even more obscure than the unidentified structures, and so they provide a stringent test of our algorithms' ability to isolate wave-like patterns from noisy data. 

\begin{figure}
\resizebox{6in}{!}{\includegraphics{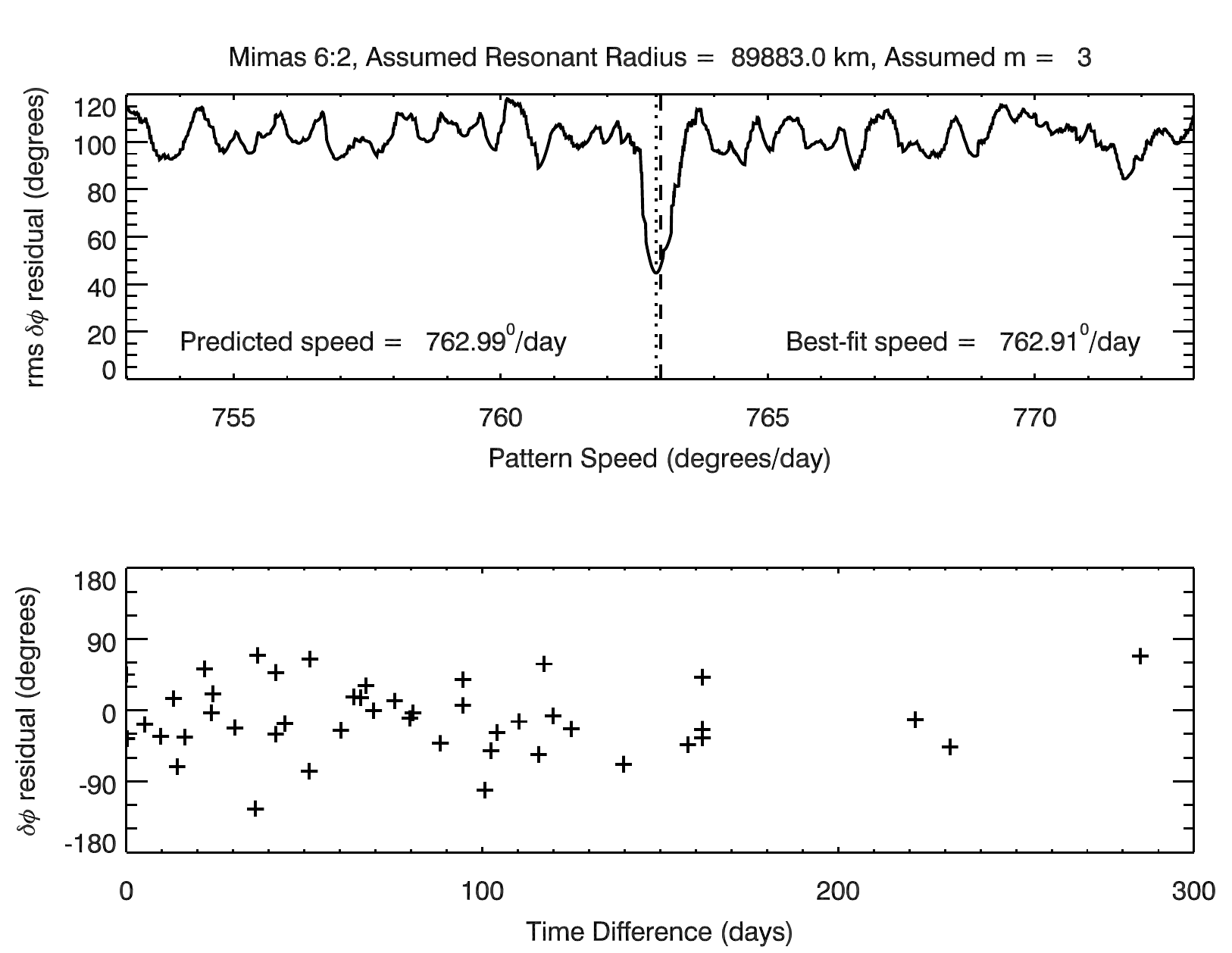}}
\caption{Results of a wavelet analysis of the wave near the Mimas 6:2 resonance which considered the radial range of 89,880-89,900 km and a wavelength range of 0.5 to 5 km. The top panel shows the $rms$ phase difference residuals as a function of pattern speed, assuming the wave is an $m=+3$ pattern. The dashed line marks the expected pattern speed for the Mimas 6:2 inner Lindblad resonance ($e^3$ type). There is a clear minimum in the residuals very close to the predicted location. The bottom panel shows the phase difference residuals (observed-expected) as a function of time difference between the observations for this best-fit solution.}
\label{Mim62res}
\end{figure}

\begin{figure}
\resizebox{6in}{!}{\includegraphics{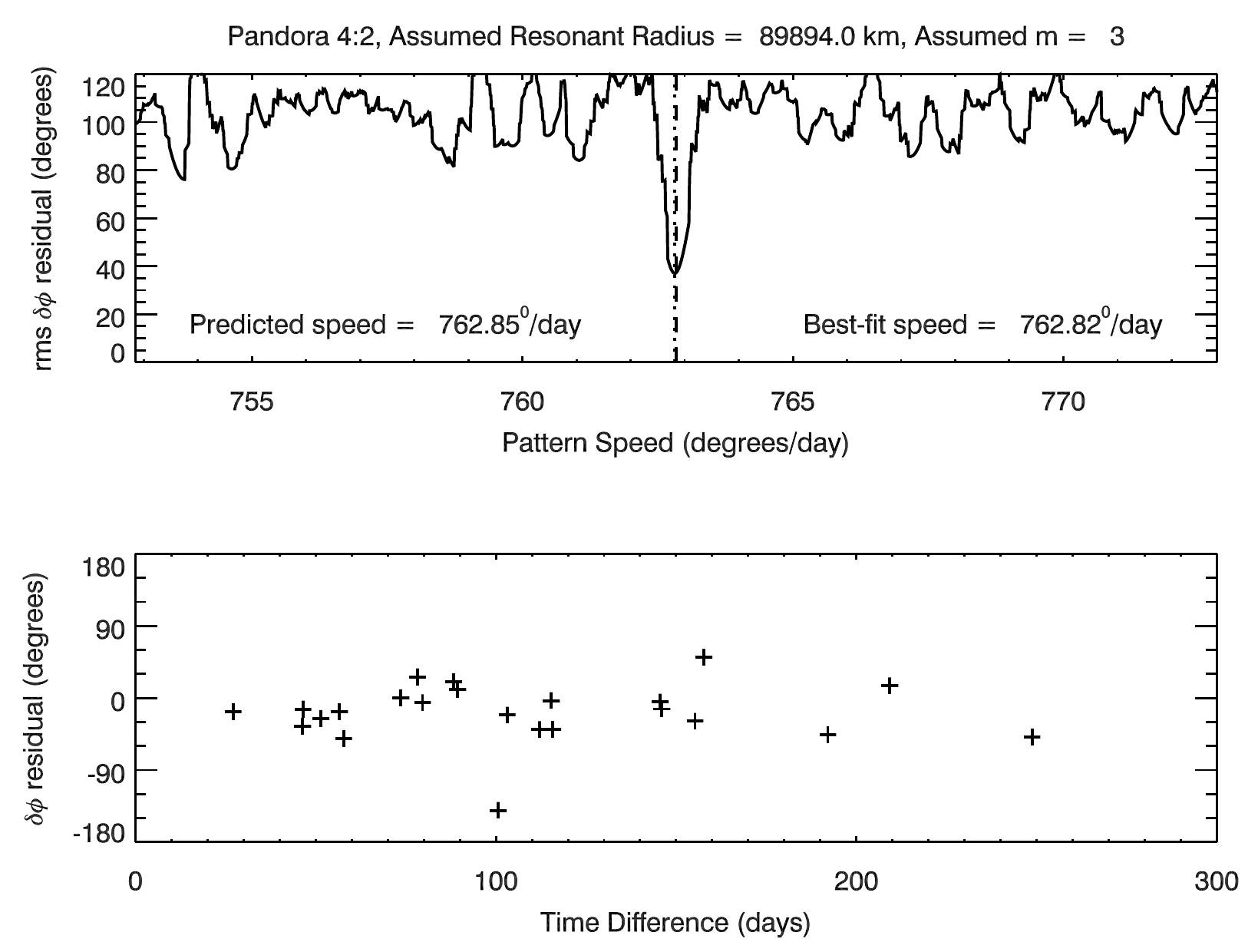}}
\caption{Results of a wavelet analysis of the wave near the Pandora 4:2 resonance which considered the radial range of 89,900-89,920 km and a wavelength range of 0.5 to 5 km. The top panel shows the $rms$ phase difference residuals as a function of pattern speed, assuming the wave is an $m=+3$ pattern. The dashed line marks the expected
pattern speed for the Pandora 4:2 inner Lindblad resonance. There is a clear minimum in the residuals very close to the predicted location. The bottom panel shows the phase difference residuals (observed-expected) as a function of time difference between the observations for this best-fit solution.}
\label{Prom42res}
\end{figure}

Figures~\ref{Mim62res} and~\ref{Prom42res} show the results of our wavelet analysis for both of these waves.  We find that each wave does indeed yield a minimum in the $rms$ phase difference residuals near the expected pattern speed when we assume that $m=+3$.\footnote{Note that since these are not first-order waves, the pattern speed does not equal the moon's orbital speed. Instead we have for the Pandora 4:2 resonance $\Omega_P=(4n_{Pandora}-\dot\varpi_{Pandora})/3=762.852^\circ$/day, while for the Mimas 6:2 resonance $\Omega_P=(6n_{Mimas}-3\dot\varpi_{Mimas})/3=762.988^\circ$/day.} The $rms$ dispersions of the residuals for these best-fit models are not as good as they are for the waves considered by \citet{HN13}, most likely because of their relatively low signal-to-noise. Even so, the minima are clear, and there are no similar minima near the expected pattern speeds for other values of $m$.  Thus our methods do seem to be able to identify the correct $m$-numbers and pattern speeds of waves even in the presence of a noisy background. Furthermore, we can now confirm the tentative wave identifications made by \citet{Baillie11}.

\begin{figure*}
\centerline{\resizebox{5in}{!}{\includegraphics{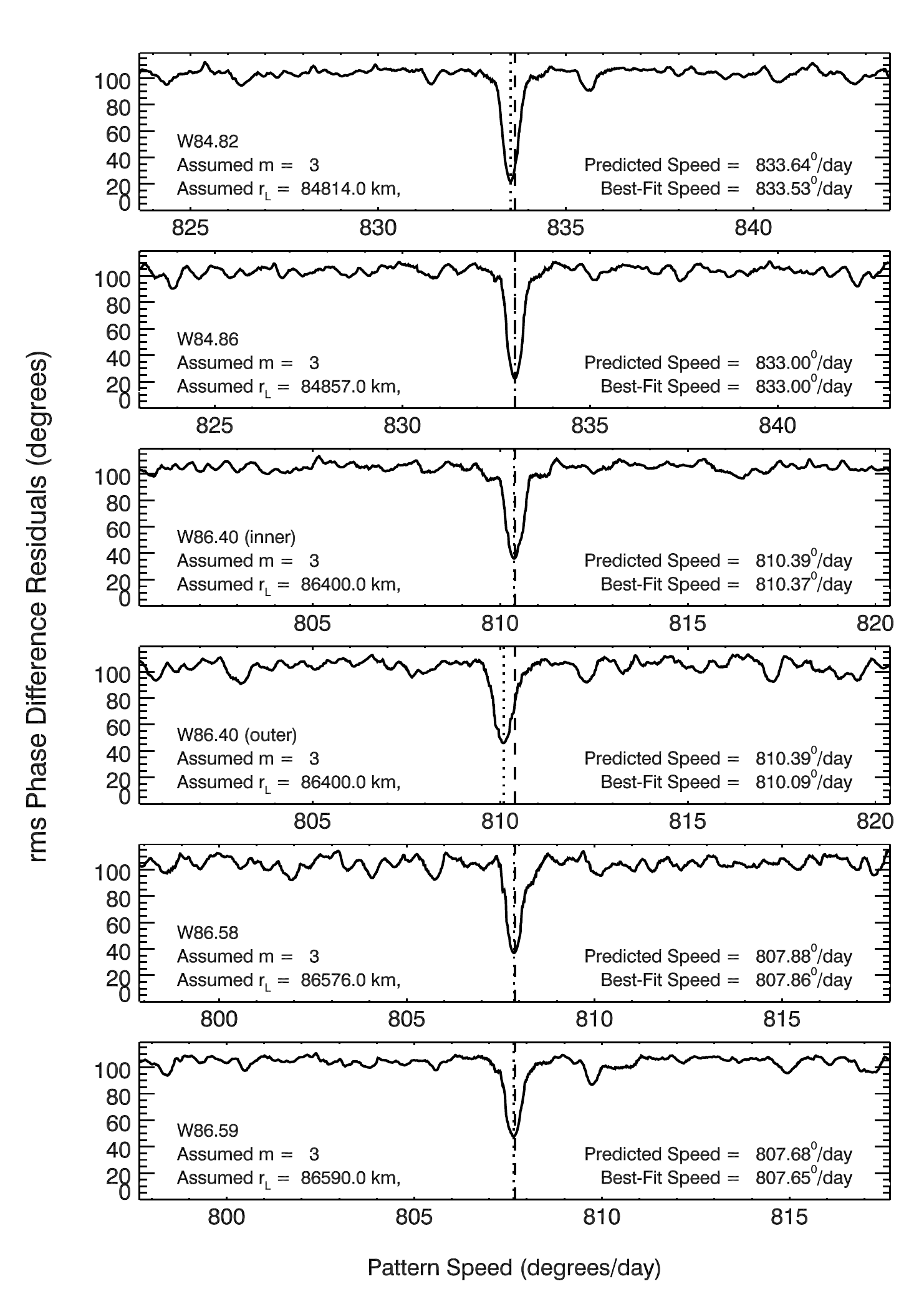}}}
\caption{Plots showing the $rms$ phase difference residuals as a function of pattern speed for the unidentified waves within plateaux P5 and P7, assuming $m=+3$ and using a wavelength range of 1-5 km for all cases except for the outer part of W86.40, which used a wavelength range of 0.3-1 km. The dashed lines mark the predicted pattern speed for each wave at the resonant location provided by \citet{Baillie11}, while the dotted lines are the pattern speeds that give the minimum variance in the residuals. \citet{Baillie11} did not provide a resonance location for W86.59, and so here the predicted resonance location has been chosen to match the observed pattern speed.}
\label{patspeed3}
\end{figure*}

\begin{figure*}
\centerline{\resizebox{5in}{!}{\includegraphics{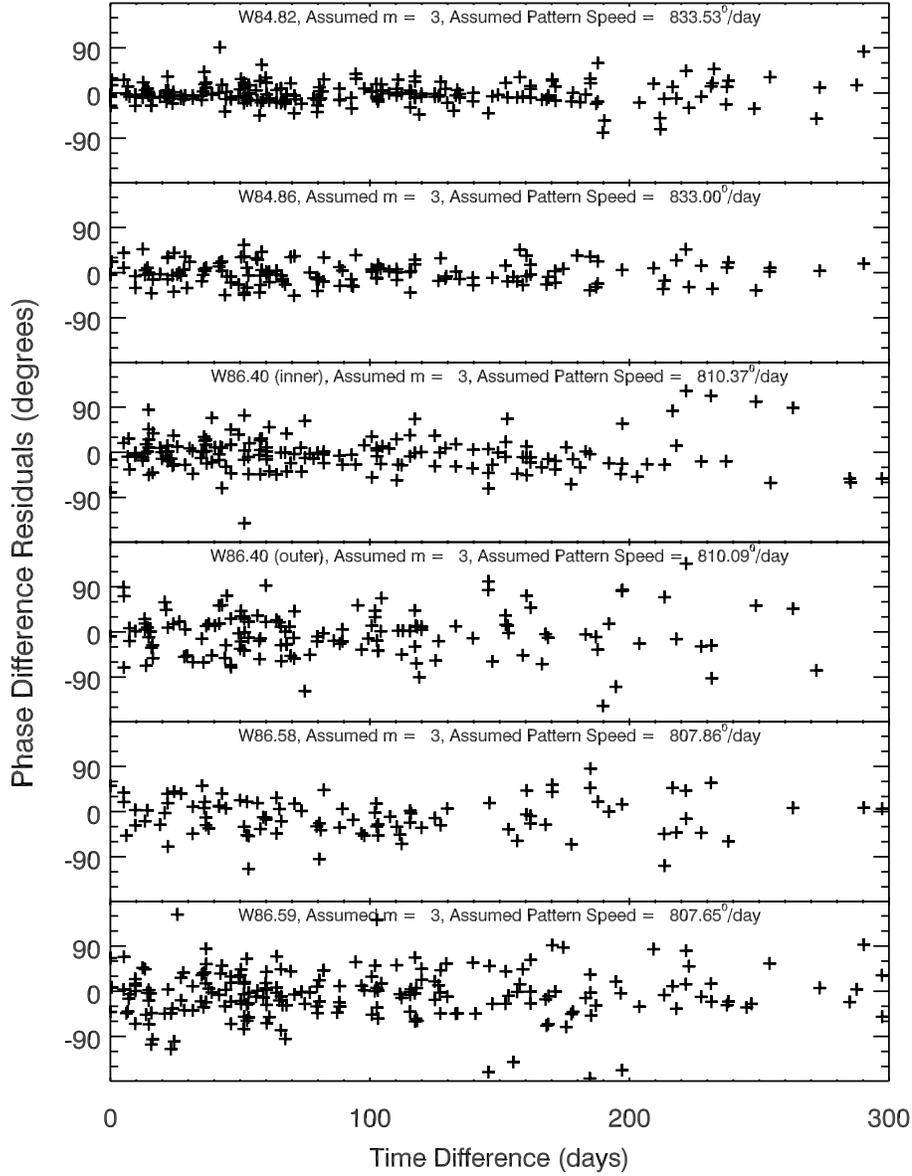}}}
\caption{Plots showing phase difference residuals (observed-predicted) for each of the  waves, assuming each pattern has the indicated $m$-number and pattern speed, which correspond to the best-fit values shown in Figure~\ref{patspeed3}.}
\label{residual3}
\end{figure*}

\begin{figure*}
\centerline{\resizebox{6.5in}{!}{\includegraphics{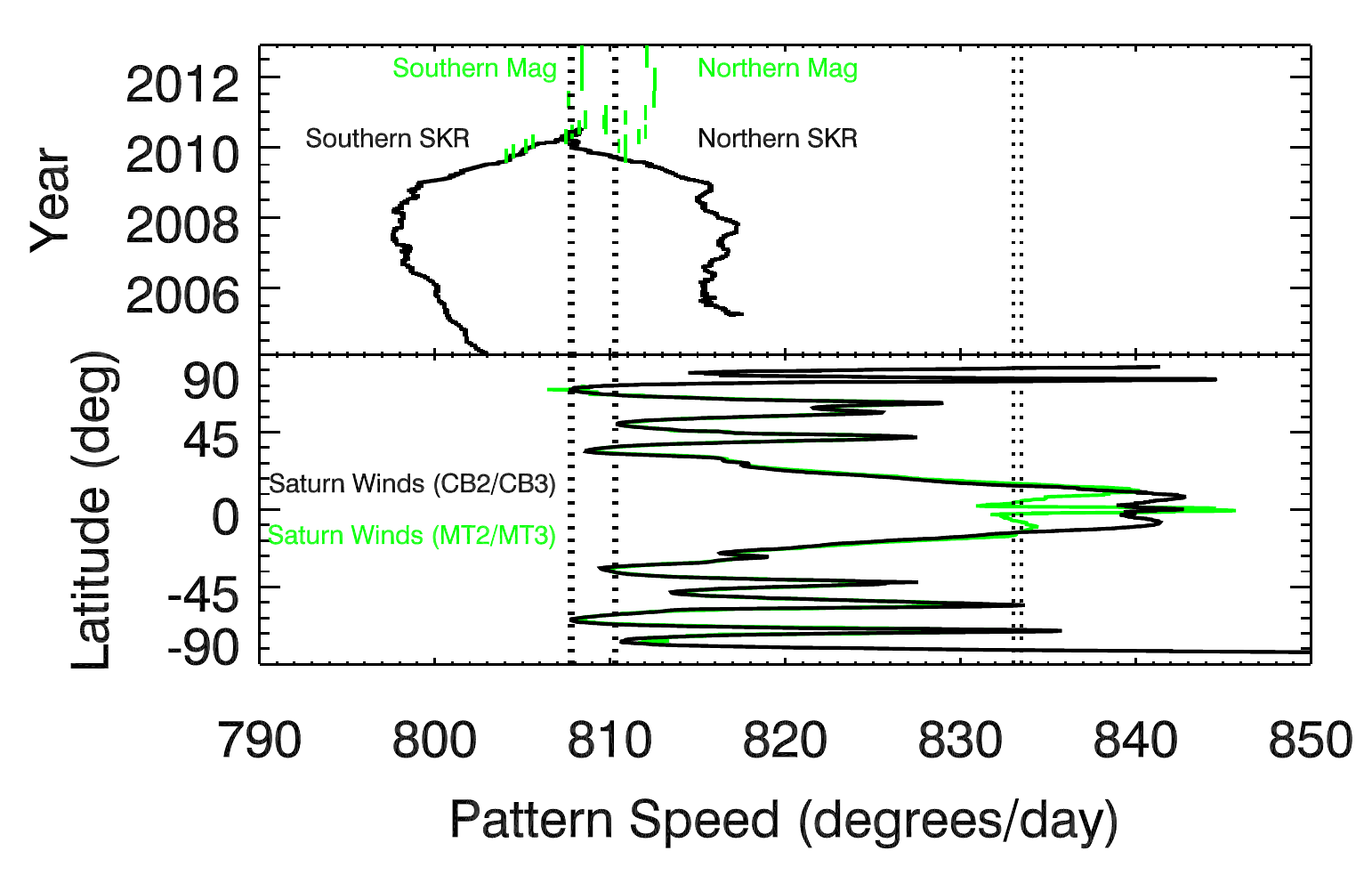}}}
\caption{Comparing the pattern speeds of the $m=+3$ waves with the rotation rate of Saturn. In both panels above, the dotted lines mark the pattern speeds of the five $m=+3$ waves. The top panel compares these pattern speeds with the estimates of the SKR rotation rates during the first 7 years of the Cassini Mission  \citep{Lamy11}, and the planetary period oscillations in the magnetic field \citep{Provan13}. The bottom panel shows  the rotation rates of Saturn's winds as a function of latitude from \citet{gm11}. Note the rotation rates of Saturn's equatorial jet depends on whether the images examined were obtained in continuum (black) or methane-band (green) filters. }
\label{satper}
\end{figure*}

\begin{table}
\caption{Pattern Speeds of selected atmospheric phenomena, modified from \citet{cp13}.}
\label{satpertab}
\resizebox{6.5in}{!}{\begin{tabular}{|c c c c|} \hline
Phenomena & Rotation Period & Pattern speed & Reference \\ \hline
W84.82 & 10.365 h & 833.5$^\circ$/day & This work \\
W84.86 & 10.372 h & 833.0$^\circ$/day & This work  \\
Estimate of bulk rotation from occultations & 10.543 h & 819.5$^\circ$/day & \citet{AS07} \\
Estimate of bulk rotation from potential vorticity & 10.570 h & 817.4$^\circ$/day & \citet{Read09} \\
IAU System III & 10.656 h & 810.8$^\circ$/day & \citet{Davies83} \\
North Polar Hexagon & 10.656 h & 810.8$^\circ$/day & \citet{slv14} \\
W86.40 & 10.662 h & 810.4$^\circ$/day & This work \\
Great White Spot Vortex & 10.667 h & 810.0$^\circ$/day & \citet{Sayanagi13} \\
String of Pearls & 10.686 h & 808.5$^\circ$/day & \citet{Sayanagi14} \\
Great White Spot Head & 10.693 h & 808.0$^\circ$/day & \citet{Sayanagi13} \\
W86.58 & 10.695 h & 807.9$^\circ$/day & This work \\
W86.59 & 10.698 h & 807.7$^\circ$/day & This work \\
\hline
\end{tabular}}
\end{table}

Turning to the unidentified waves (W84.82, W84.86, W86.40, W86.58 and W86.59), we were surprised to find that all five of these patterns also showed clear minima when we assumed $m=+3$. As shown in Figures~\ref{patspeed3} and~\ref{residual3}, the best-fit models for these waves typically have smaller residuals than those derived above for the Mimas 6:2 and Pandora 4:2 waves. This probably reflects the somewhat higher visibility of these waves in the individual profiles.

Waves W84.82 and W84.86 both inhabit the same plateau P5, and are only 35 kilometers apart (See Figure~\ref{overview}).   We computed the phase differences for these waves using the radial ranges 84810-84830 km and 84850-84880 km. We only considered wavelengths between 1 and 5 km for these waves in an effort to minimize any contamination from fine-scale stochastic structure in this region. Figure~\ref{patspeed3} shows that there is a clear minimum in the $rms$ residuals for W84.82 when $m=+3$ and $\Omega_p=833.53^\circ/$day, or just 0.12$^\circ$/day slower than the expected pattern speed using the resonant radius of 84,814 km provided by \citet{Baillie11}, which is actually the predicted radius for the nearby 4:2 Pan resonance. This slight shift would imply the actual resonant location is 8 km exterior to this reported position, or around 84,822 km. This number is not unreasonable given that  the prominent density variations start at about this location (see Figure~\ref{m3prof}), and provides further evidence that this wave is not generated by the Pan resonance. Wave W84.86, on the other hand, shows a minimum for $m=+3$ and $\Omega_p=833.00^\circ/$day, which perfectly matches the expected pattern speed using the resonant radius of 84,857 km provided by \citet{Baillie11}. Note that the difference between these two pattern speeds is significant even if we consider only time separations up through 300 days. Thus we have two waves with the same symmetry properties and very similar, but not identical, pattern speeds.

Turning to wave W86.40, which is located in plateau P7, we first note that this wave seems to extend further from the putative resonance than W84.82 and W84.86 do. Furthermore, many profiles show hints of strong sub-kilometer wavelength structures 30-50 km exterior to the resonance. The wavelet analysis of \citet{Baillie11} indicates that all these variations could potentially be ascribed to a single extensive wave, but to be sure, we decided to analyze the two parts of the wave separately, using radial ranges of 86400-86420 km and 86430-86450 km for the inner and outer part, respectively. We continued to use the wavelength range of 1-5 km for the inner part of the wave, but we used 0.1 to 3 km for the outer part in order to capture the finer-scale structure further from the resonance. As shown in Figure~\ref{patspeed3}, the inner part of the wave yields a clear $m=+3$ pattern with a pattern speed of 810.37$^\circ$/day, which matches the expected value for the resonant radius of 86,400 km derived by \citet{Baillie11}. Intriguingly, the outer part of the wave also shows a minimum when we assume $m=+3$, but at a slightly slower pattern speed of 810.09$^\circ$/day (this difference in pattern speed remains even if we use the same wavelength range for both parts of the wave). The residuals for the outer part of the wave have a larger scatter, which probably reflects the shorter wavelength  of the pattern in this region. The slightly slower best-fit pattern speed, which corresponds to a resonant radius of 86,418 km, could just represent systematic errors in the analysis due to the short wavelengths involved, but it could potentially indicate that there are actually two overprinted $m=+3$ waves occupying this region. 

\citet{Baillie11} identified three additional features (29-31) in the outer part of plateau P7. When we inspected the VIMS profiles, we could only discern two waves, which we designate W86.58 and W86.59. \citet{Baillie11} provided a resonance location of 86,576 km for W86.58, but did not provide a resonant radius for W86.59. The latter feature overlaps the outer edge of the plateau (see Figure~\ref{m3prof}), which would complicate the interpretation of any wavelength trends. When we analyzed W86.58 using a radial range of 86,575-86,585 km and a wavelength range of 1-5 km, we obtained a reasonably clear minimum in the phase difference residuals with $m=+3$ and a pattern speed close to the predicted rate of 807.88$^\circ$/day (see Figure~\ref{patspeed3}). For W86.59, we considered a radial range of 86,595-86,605 km and a wavelength range of 1-5 km, but the sharp background slope contaminated the phase measurements. We therefore applied a high-pass filter to these data  by subtracting a copy of the profile smoothed by a boxcar average with length 2 km. Phase differences derived from these filtered light curves yielded a clear minimum in the phase difference residual with $m=+3$ and a pattern speed of 807.65$^\circ$/day, consistent with the location of this wave. Note the $rms$ scatter in the phase difference residuals are larger for these two waves than they are for W84.82 and W84.86 (see Figure~\ref{residual3}), most likely because of their lower signal-to-noise. 

It is remarkable that we have at least five  $m=+3$ waves with pattern speeds between 807$^\circ$/day and 834$^\circ$/day. These pattern speeds do not correspond to the oscillation frequencies of planetary normal modes such as those computed by \citet{MarleyPorco93} or \citet{Marley14}. Instead, they are close to Saturn's rotation rate. Saturn's magnetic dipole is almost perfectly aligned with its rotation axis, so its internal rotation rate is still uncertain, with various indirect estimates given rotation rates between 817$^\circ$/day and 820$^\circ$/day \citep{AS07, Read09}.  On the other hand, the clouds in Saturn's atmosphere rotate around the planet at rates ranging between $808^\circ$/day and $828^\circ$/day at mid-latitudes, and up to 850$^\circ/$day in the equatorial jet \citep{gm11, slv14}, while various periodic phenomena in Saturn's magnetosphere like the Saturn Kilometric Radiation (SKR) exhibit at least two components with frequencies ranging between 795$^\circ$/day and 820$^\circ$/day over the last few years \citep{Ye10, Lamy11, Andrews12, Provan13}.  Table~\ref{satpertab} compares our observed pattern speeds for the $m=+3$ waves with a number of published estimates of rotation rates for various features in the planet's atmosphere, while Figure~\ref{satper} compares our pattern speeds with both the variable frequencies of the magnetospheric phenomena and Saturn's winds. Intriguingly, the pattern speeds of waves W86.40, W86.58 and W86.59 are close to the magnetospheric frequencies and Saturn's westward jets, while the pattern speeds of W84.82 and W84.86 are close to the rotation rate of Saturn's eastward jets.

Given the similarity between these wave's pattern speeds and Saturn's atmospheric and magnetospheric rotation rates, it seems likely that all of these waves are generated by ``tesseral'' resonances with structures rotating with the planet. Such resonances were first suggested by \citet{Franklin82} but turned out to be undetectable in the Voyager data  \citep{Holberg82}. The waves visible in the Cassini data all have $m=+3$, and occur where the Keplerian mean motion is around 3/2 Saturn's rotation rate,  so they would all represent 3:2 tesseral resonances. This is sensible, because in the C ring only $m=+3$ resonances have pattern speeds that can match the planet's rotation rate (see Figure~\ref{reslocs}). However,  different tesseral resonances could occur in other parts of the rings. Indeed,  there are several patterns in both the D ring and Roche Division which appear to be generated by resonances with multiple periodic perturbations with effective pattern speeds around 800$^\circ$/day and 820$^\circ$/day \citep{Hedman09}, and these may correspond to the 2:1 and 3:4 tesseral resonances, respectively. In addition, the dusty spokes that form close to the planetary co-rotation radius also exhibit periodicities that have been tied to Saturn's rotation rate \citep{PD82, Porco83, Mitchell13}. However, all of these previously-studied structures involve ring material composed primarily of micron-sized particles. These tiny grains are  very sensitive to non-gravitational forces and thus might be influenced by the same electromagnetic phenomena that modulate the SKR. The C-ring waves, by contrast, involve much denser rings composed of pebble-to-boulder sized ice particles, and thus are almost certainly generated by periodic gravitational perturbations. These waves presumably trace long-lived gravitational anomalies inside Saturn that are carried around the planet by winds moving at slightly different rates. 

In principle, the masses associated with these anomalies can be estimated from the waves' amplitudes. In practice, precise mass estimates would require detailed measurements of the wave profiles and assumptions about the spatial form of the anomaly, and such analyses are beyond the scope of this report. However, we can derive order-of-magnitude estimates of the perturbing masses by simply noting that the amplitudes of the relevant waves  are comparable to those generated by the Mimas 6:2 and Pandora 4:2 resonances. The perturbations applied to the rings by these two satellite resonances are proportional to $Me^p$, where $M$ is the moon's mass, $e$ is its orbital eccentricity, and  $p=3$ for the Mimas 6:2 resonance while $p=1$ for the Pandora 4:2 resonance. Numerically, these factors are $3\times10^{14}$ kg and $6\times10^{14}$ kg for the Mimas 6:2 and Pandora 4:2 resonances, respectively. If the unidentified $m=+3$ waves are generated by compact mass anomalies inside the planet, those masses would need to be of the same order as these factors to produce waves with similar amplitudes (neglecting various coefficients of order unity). Hence each wave would require a mass anomaly of order $10^{14}-10^{15}$ kg to generate the observed waves. This is comparable to the mass of a kilometer-sized icy satellite, and corresponds to a very small perturbation in the planet's density 
\begin{equation}
\frac{\delta\rho}{\rho_d} \simeq 10^{-11}\left(\frac{\rm 0.7 g/cm^3 }{\rho_d}\right)\left(\frac{R_s^3}{V_d}\right) 
\end{equation}
where $\rho_d$ and $V_d$ are the mean mass density and volume of the region in the planet responsible for the mass anomaly. If the anomaly corresponds to a large-scale $m=3$ perturbation in the planet's envelope, then $V_d$ will be of order $R_s^3$ and the density perturbation may be only one part in $10^{11}$. However, if the anomalies are associated with something comparable in scale to Saturn's storm clouds  ($\sim$ 2000 km, Dyudina {\it et al.} 2007), then the density contrast could be more like $10^{-6}$.  
\nocite{Dyudina07}

Unless these anomalies have a pure $m=3$ structure inside the planet, we would expect additional gravitational tesseral resonances to be found elsewhere in the main rings. Thus far, no waves generated by tesseral resonances have been found outside the C ring, but this may be because the 3:2 tesseral resonances are the ones that lie closest to the planet (excluding the 2:1 resonances, which fall in the D ring), and so produce the strongest perturbations on the ring material. However, careful searches for unidentified waves elsewhere in the B and A rings may eventually reveal additional examples.

\begin{figure}
\resizebox{6in}{!}{\includegraphics{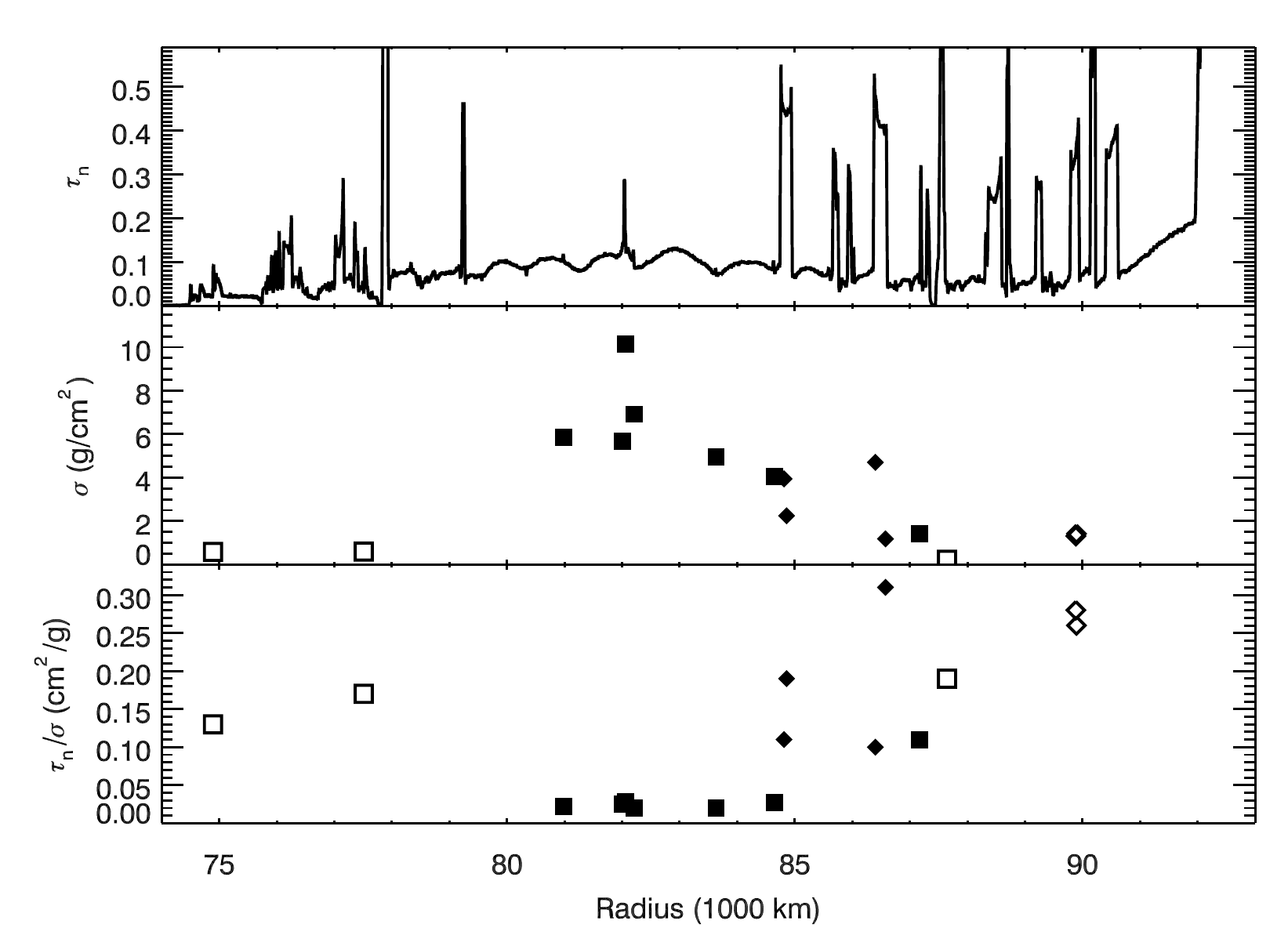}}
\caption{Comparisons of the C-ring's normal optical depth and surface mass density. The top panel shows an optical depth profile derived from the Rev 89 $\gamma$Crucis occultation. The center and bottom panels show the surface mass density and opacity of the ring derived from various density waves. The open symbols are values derived by \citet{Baillie11}, while filled symbols utilize the $m$-values derived in this work and \citet{HN13}.  Squares represent waves in the background C ring, while diamonds correspond to measurements in plateaux. Note that the ring's surface mass  reaches a maximum and the opacity is at a minimum in the middle of the C ring.}
\label{sigma}
\end{figure}

Regardless of the source of these waves, their identification as $m=+3$ patterns exacerbates a pre-existing incongruity in the estimated mass densities of the plateaux. Again, using the estimates of $\sigma_0/|m-1|$ and $\tau_n/\sigma_0|m-1|$ from \citet{Baillie11}, we find surface mass densities  of 3.94, 2.24, 1.18 and 4.70 g/cm$^2$ and opacities of 0.11, 0.19, 0.10, and 0.31 cm$^2$/g for W84.82, W84.86, W86.40 and W86.58 respectively (see Table~\ref{sumtab}; as mentioned above, \citet{Baillie11} did not provide mass density estimates for W86.59). These mass density estimates are surprising because they are slightly less than those of the middle C ring, even though the normal optical depths of these plateaux are several times larger (see Figure~\ref{sigma}).  This indicates that the substantially larger optical depths found in the plateaux do not correspond to a similarly elevated mass density. Instead, it appears that the mass density of the C ring peaks at around 82,000 km, and the plateaux have a comparable mass density to nearby lower-optical depth portions of the ring. Indeed, the nearby waves W84.64 (in the background ring) and W84.82 (on a plateau) yield nearly the same mass density, even though their optical depths differ by a factor of four.

Another way to look at these variations is to consider the opacity parameter $\tau_n/\sigma_0$, which is inversely proportional to the effective mean particle size and average particle density. As shown in Figure~\ref{sigma} and Table~\ref{sumtab}, the C-ring's opacity is at a minimum in the middle C ring, and is substantially larger within the plateaux. In particular, the opacity from W84.82 is roughly four times larger than W84.64. Such rapid variations in the ring's opacity indicate that the average sizes and/or internal densities of particles in the plateaux differ dramatically from those in the rest of the C ring. Data from Cassini radio occultations do indicate that there are substantial variations in the particle size distribution across the C ring, but the trends found by those experiments only seem to further confuse the situation. Simultaneous measurements made at multiple radio wavelengths reveal that the background C-ring has a 30\% higher optical depth to 3.6 cm radiation than it does to  13.0 cm radiation, while this difference is much reduced in the plateaux \citep{Cuzzi09}. This suggests that the middle C ring has a larger fraction of centimeter-sized particles than the plateaux do, which implies that the particles in the plateaux are somewhat {\it larger} on average than the particles elsewhere in the C ring.\footnote{Note tht microwave measurements are sensitive primarily to particles in the centimeter to
decameter size range.}  This is problematic because larger particles have lower surface-area-to-volume ratios and so we should expect regions with larger typical particle sizes to have lower opacities, which is exactly the opposite of what we observe. Perhaps there are significant
differences in the ring-particles' composition and/or internal mass density between these two regions. Ongoing analyses of Cassini occultation measurements, which indicate that there are also differences in the upper end of the particle size distributions between the plateaux and the background ring (Colwell et al. 2011, in prep), may help to clarify this situation. \nocite{Colwell11}

\pagebreak

\section{W85.67, a migrating m=-1 wave?}
\label{W85.67}

\begin{figure}
\centerline{\resizebox{6in}{!}{\includegraphics{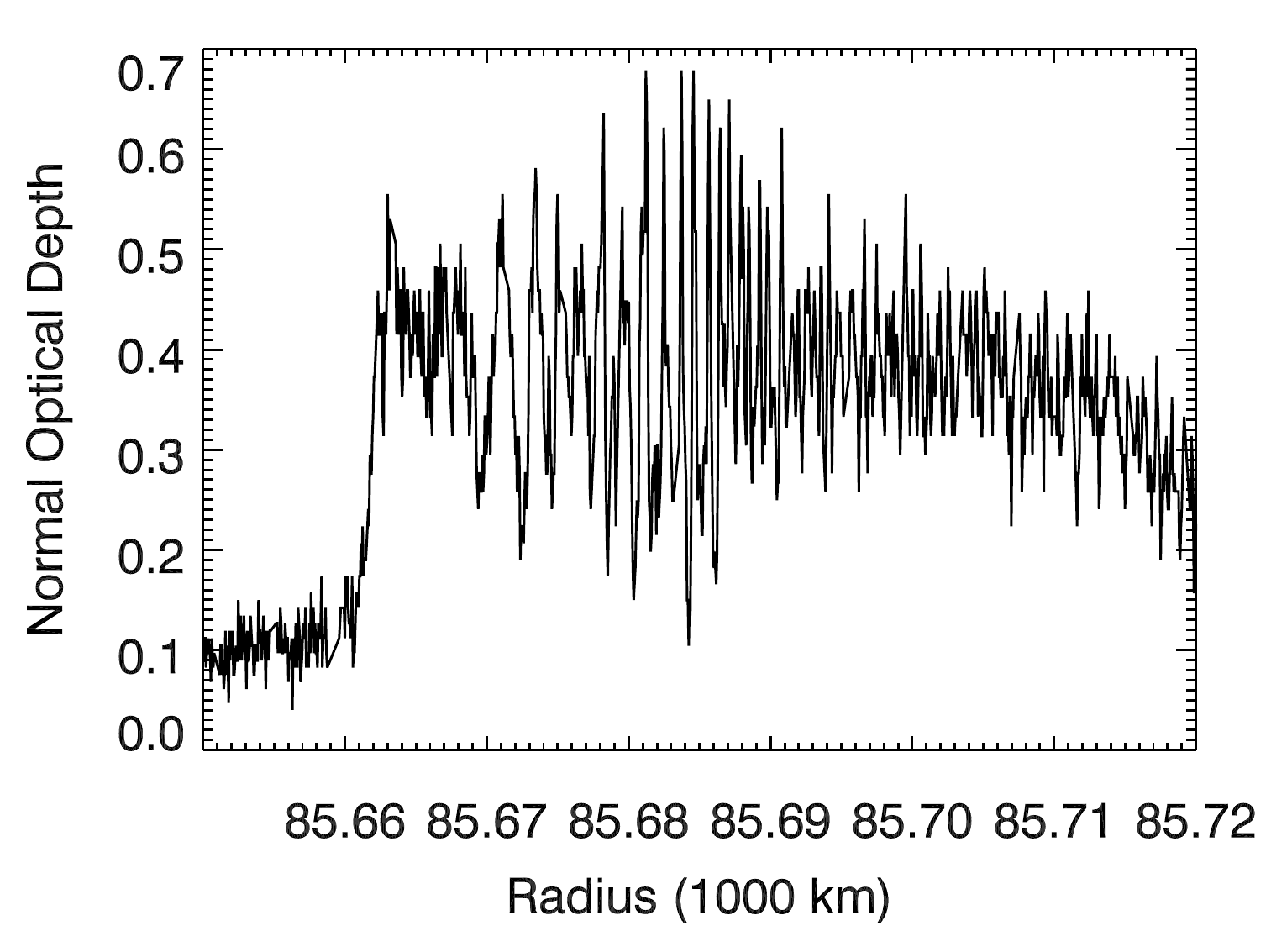}}}
\caption{Sample profile of the W85.67 wave from the Rev 106 RCas occultation (B=56.04$^\circ$). Note that
this wave occupies a 100-km wide plateau known as P6, and  its wavelength clearly decreases with increasing radius.}
\label{W85.67prof}
\end{figure}

\begin{figure}
\resizebox{6in}{!}{\includegraphics{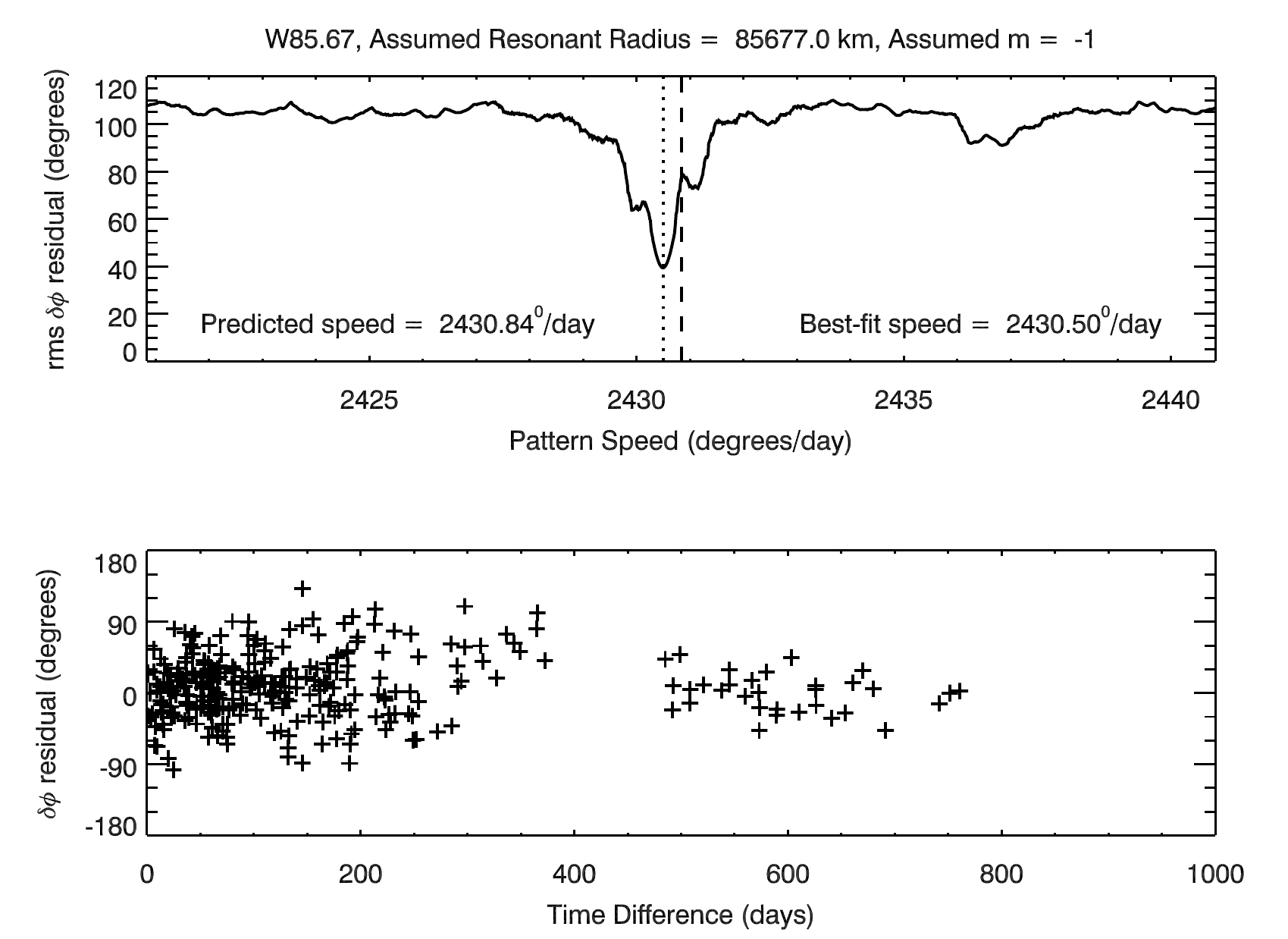}}
\caption{Results of a wavelet analysis of wave W85.67 which considered the radial range of 85,675-85,690 km and a wavelength range of 1 to 5 km. The top panel shows the $rms$ phase difference residuals as a function of the assumed pattern speed, assuming the wave is an $m=-1$ pattern. The dashed line marks the expected
pattern speed for such a structure with the resonant radius estimated by \citep{Baillie11}. There is a clear minimum in the residuals very close to the predicted location. The bottom panel shows the phase difference residuals (observed-expected) as a function of time difference between the observations for the best-fit solution.}
\label{W85.67res}
\end{figure}

\begin{figure}
\centerline{\resizebox{6in}{!}{\includegraphics{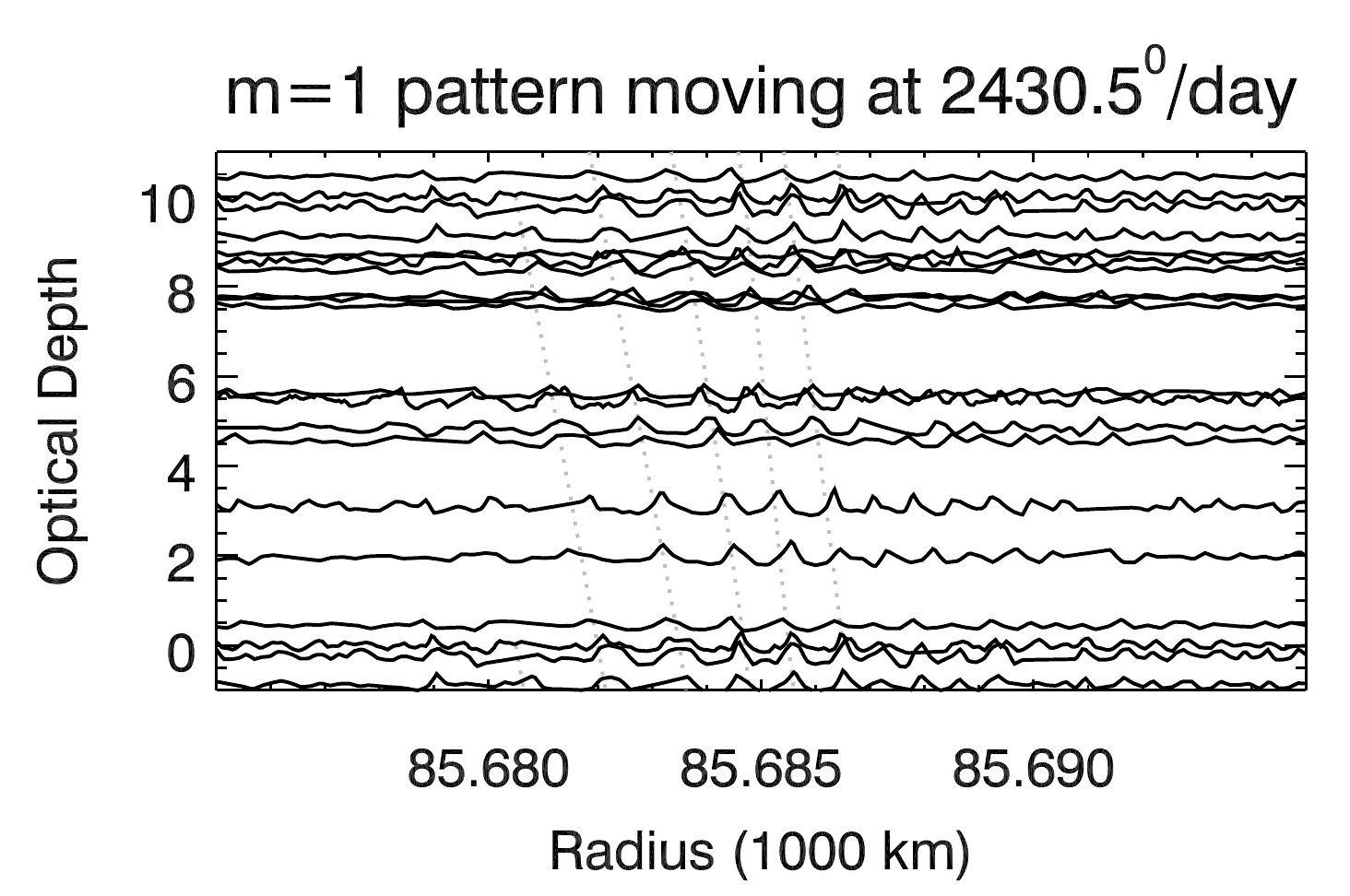}}}
\caption{Selected profiles of the W85.67 wave versus radius, with vertical offsets proportional to the estimated phase of the pattern assuming $|m|=1$ and $\Omega_P=2430.5^\circ$/day. Note that the same set of profiles are repeated  near the top and bottom of the panel in order to make the pattern easier to see. Together, the profiles are consistent with a spiral pattern where the opacity maxima shift to smaller radii as the phase increases following the dotted lines (included to guide the eye), as expected for a trailing wave. Also, there does appear to be a single arm, as expected for an $|m|=1$ pattern. However, the wavelength of the pattern clearly gets smaller with increasing radius, which implies that it is propagating outwards, which is inconsistent with its fast pattern speed.}
\label{W85.67phase}
\end{figure}


The final, and perhaps most perplexing,  wave we  will consider here was designated W85.67 by \citet{Colwell09} (also known as wave $d$ in \citet{Rosen91} and feature 27 in \citet{Baillie11}).  This wave lies between the W84.86 and W86.40 waves discussed above, and also occupies a plateau (P6, see Figure~\ref{overview}). Compared with the $m=+3$ waves, however, this feature  has much more prominent and obvious opacity variations (see Figure~\ref{W85.67prof}). Its wavelength appears to decrease with increasing radius, which suggests that this is also an outward-propagating wave with a positive $m$-number. Table~\ref{obstab} lists the occultation profiles considered in this study. Since we must again contend with elevated levels of stochastic fine-scale structure in this region, we consider only wavelengths between 1 and 5 km, but our results are insensitive to the exact range.

We searched for patterns with $m=+1$ through $m=+15$ but did not find a clear minimum in the $rms$ phase difference residuals within 10$^\circ$/day of any of the expected pattern speeds. Out of desperation, we considered negative values of $m$\footnote{We also considered $m=0$, which corresponds to a pattern with no azimuthal variations but a temporal oscillation frequency equal to the local radial epicyclic frequency. This case also failed to yield a sensible solution.}, and to our surprise found a strong minimum near the expected pattern speed when we used $m=-1$. This solution is shown in Figure~\ref{W85.67res}, which considers phase differences from occultations up to 1000 days apart in order to demonstrate that this solution is consistent with a steady pattern speed of 2430.5$^\circ$/day. We note that the dispersion around this best fit solution is rather large compared to the other unknown waves, but still all the  phase difference residuals are within $\pm$90$^\circ$ of zero, even with time separations approaching 1000 days. 

An $m=-1$ solution for this wave  is puzzling because the $\Omega_P$ implies an outer Lindblad resonance driven by a perturbation period of order 3 hours, while the morphology of the wave is similar to outward-propagating waves driven by inner Lindblad resonances with moons.  Furthermore, the derived pattern speed is  actually somewhat slower than we might have expected. If this is a density wave with a resonant radius of 85,677 km (as derived by Baillie et al. 2011) we would instead predict a pattern speed of  2430.85$^\circ$/day, about 0.35$^\circ$/day faster than the observed value. To match the observed pattern speed, the resonant radius would need to be around 85,685 km, which lies within the wave itself. Such a discrepancy between the inferred resonant location of the wave has not been observed in any of the other waves we have examined, and reinforces the idea that something is odd about this wave.

Given these unusual observations, we decided to check our result by looking at the profiles themselves. Figure~\ref{W85.67phase} shows selected profiles, sorted by the predicted phase $\phi$ if we assume an $|m|=1$ pattern rotating at 2430.5$^\circ$/day. We clearly see the spiral pattern where the peaks in optical depth shift systematically inwards with increasing phase, as expected for a trailing spiral. Thus the best-fit $m=-1$ solution does indeed organize the data sensibly.

Probably the simplest way to reconcile W85.67's morphology and pattern speed is for it to be a bending wave instead of a density wave. Bending waves propagate in the opposite direction from density waves, so an outward-propagating bending wave would be due to an outer vertical resonance, and thus have a higher pattern speed than the local mean motion, as we observe. However, there are major problems with identifying this feature as a bending wave.  First, we do not observe the viewing-angle dependent effects expected for bending waves. If the observed opacity variations were due to vertical structure, they should become more prominent at lower ring opening angles, and this is not observed. Also, the positions of maxima and minima should depend not only on the phase of the pattern but also the observation geometry, but if we include these effects in our analysis, it does not reduce the residuals from the best-fitting solution. In fact, with these corrections we found no useful minima for any value of $m$. Finally, the pattern speed of this wave is not appropriate for a bending wave. An $m=-1$ bending wave at this location would have a pattern speed of around $2362^\circ$/day, which is very different from the observed value, and there is no strong minimum at that pattern speed. Together, these findings strongly suggest that this feature is not a bending wave. 

\begin{figure}
\resizebox{6.5in}{!}{\includegraphics{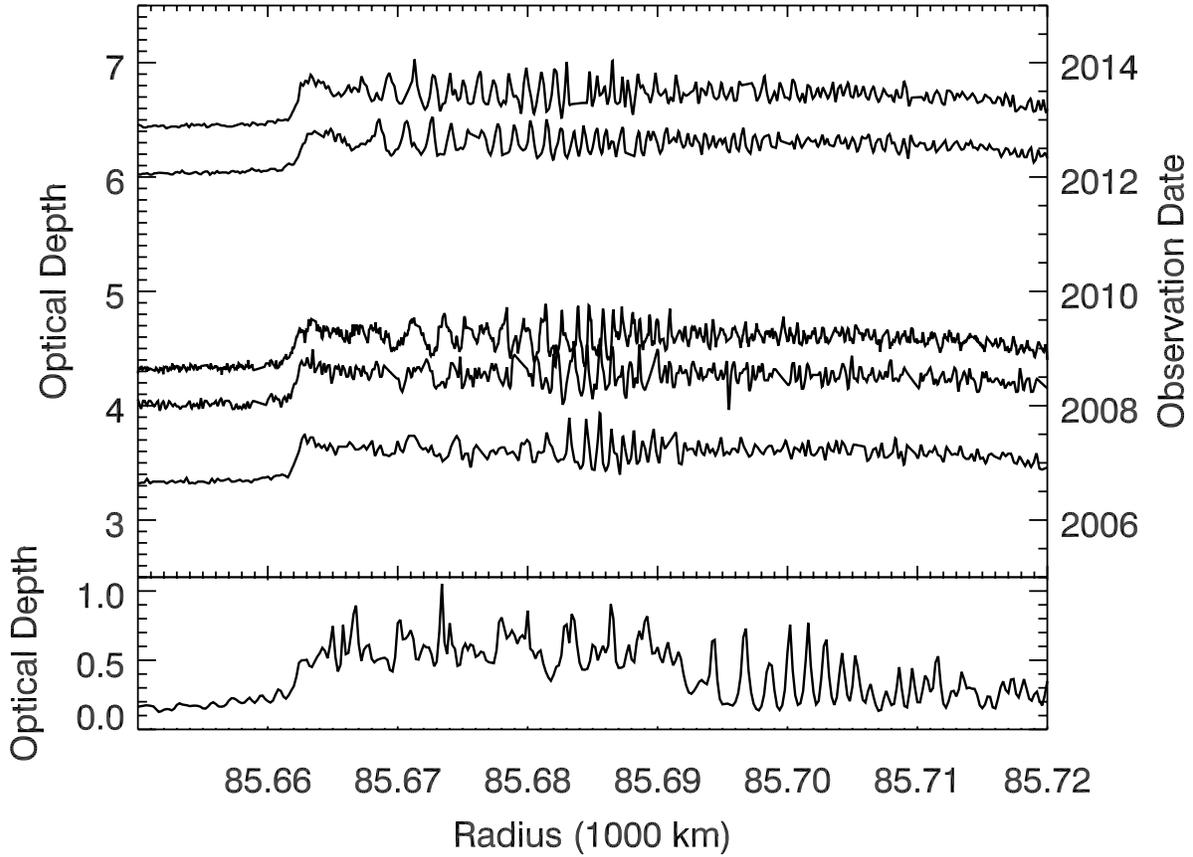}}
\caption{Profiles of W85.67 obtained over a broad range of times by Cassini and Voyager. The bottom panel show the RSS profile of this wave obtained by Voyager 1 in 1980. The top panel shows six profiles obtained by VIMS during the following occultations: Rev 41 $\alpha$ Aurigua, Rev 77 $\gamma$ Crucis, Rev 106 R Cassiopea, Rev 170 $\beta$ Pegasi, Rev 193 $\mu$ Cephii. These profiles have been vertically offset by amounts proportional to the time separation between them.  We see clearly that the wave pattern has been moving inwards over time.}
\label{W85.67mov}
\end{figure}

\begin{figure}
\resizebox{6.5in}{!}{\includegraphics{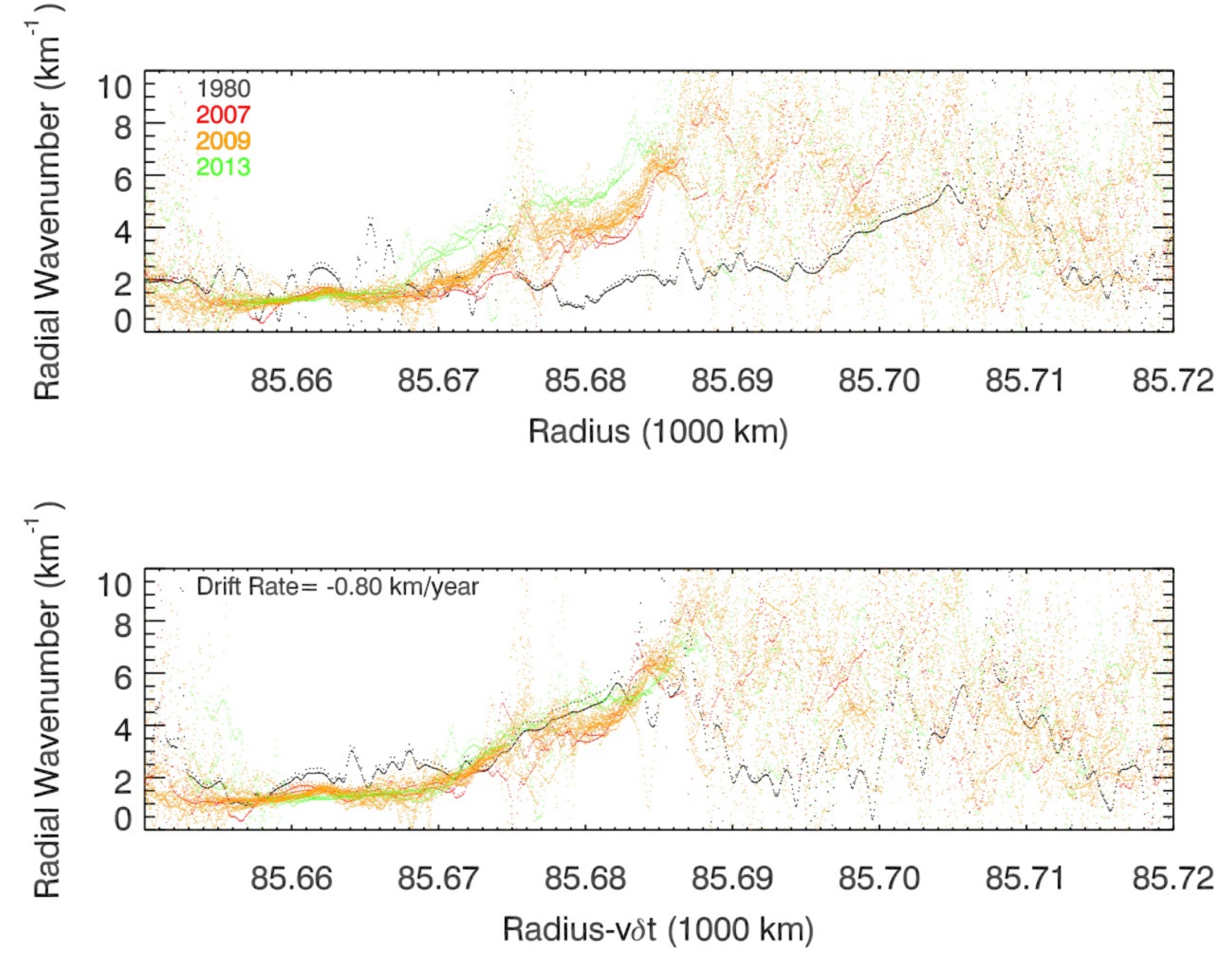}}
\caption{Plots showing the dominant radial wavenumber of the occultation profiles around W85.67, derived from wavelet analyses. The top panel shows wavenumber profiles derived from the Voyager RSS occultation in black and the Cassini VIMS occultations in various colors. Within the wave, the wavenumber increases linearly with radius, as expected, but
the location of this ramp moves steadily inwards over time. The bottom panel shows the same profiles, shifted relative to each other assuming a constant drift rate of -0.8 km/year. The wavenumber ramps all align in this case, indicating that the resonance has been moving inwards at approximately  this rate for the last 30 years.}
\label{W85.67wave}
\end{figure}

Instead, the apparent incongruity between W85.67's morphology and pattern speed could be explained by another unusual feature of this wave: it appears to be drifting slowly through the ring. Figure~\ref{W85.67mov} shows representative profiles of this wave obtained by VIMS, along with the profile derived from the Voyager Radio Science experiment. These profiles clearly demonstrate that the wave has moved moved inwards between the Voyager epoch (when it was centered around 85,700 km) and the Cassini era (where it was closer to 85,680 km). Even within the span of Cassini observations, we can see that the wave has continued to move inwards, and is now approaching the inner edge of the plateau.

A slightly different way to visualize these changes is to plot the dominant pattern wavelength in the profile as a function of radius. As in our previous wavelet analysis, we use a wavelet transform and compute the power-weighted phase of the profile as a function of radius $\phi(r)$. The radial derivative of this phase parameter $d\phi/dr$ then provides an estimate of the dominant wavenumber in the profile as a function of radius. Figure~\ref{W85.67wave} shows the wavenumber profiles for the various Cassini and Voyager occultations. In each profile, the wave appears as a nearly linear ramp in wavenumber. If we assume the wave has been moving steadily inwards at a rate of around 0.8 km/year, then the Voyager and Cassini profiles can be aligned fairly well. It therefore appears that this wave has been moving inwards at around 0.8 km/year for the last 30 years. Given the wave could be moving through regions of different surface mass densities (which also influence the wavelengths), we cannot yet provide a more precise estimate of this rate or a robust estimate of the uncertainty on this parameter. However, even this rough estimate of the wave's motion has important implications for the wave's morphology because it is comparable to the wave's group velocity.

Waves in planetary rings have a finite group velocity, so these features cannot respond instantaneously to changes in the perturbing forces.  The group velocity of a  density wave in Saturn's rings is given by the following expression \citep{Shu84}:
\begin{equation}
|v_g|=\pi G\sigma_0/\kappa
\end{equation}
where $G$ is the universal gravitational constant, $\sigma_0$ is the ring's local surface density, and $\kappa$ is the local radial epicyclic frequency, which is about 0.00025/s in this part of the C ring. The group velocity in this region is therefore $0.26(\sigma_0/1$ g/cm$^2$) km/year. This means that if the local surface mass density of this plateau is less than 3 g/cm$^2$, then the resonant radius is moving faster than the wave's group velocity. Given that the $m=3$ waves in adjacent plateaux indicate mass densities of order a few grams per square centimeter (see Figure~\ref{sigma}), this condition could indeed be met.

If the resonant radius is indeed moving radially faster than the wave itself can propagate, then the morphology of the wave can be significantly distorted. The clearest illustration of this phenomenon comes from the waves generated by the co-orbital moons Janus and Epimetheus. These two moons swap orbital positions every four years, and so the precise resonance locations move back and forth in the rings. As a result, the waves they generate exhibit a complex, time variable morphology. However, \citet{Tiscareno06} showed that the morphology of these waves can be modeled as the superposition of multiple components, each propagating at a finite speed from a particular source location. Applying the same basic concept to W85.67 yields some useful insights into this wave's structure.

 First consider a wave generated by a perturbation with a fixed period, so that the resonant location $r_L$ is fixed. In this case, at any location $r$ we are seeing a piece of the wave that was generated a time $\delta t=(r-r_L)/v_g$ before the wave was observed (note we take $v_g$ to be negative for an inward propagating wave). But what if instead the resonance location is itself moving at a speed $v_L$? If we observe the wave at a radius $r$, then we must consider both the location of the resonance now $r_{L0}$ and the location of the resonance when the wave now observed at $r$ was first generated, which we will denote $r_{L1}$. Let us now say $\delta t$ is the time it took the wave to move from $r_{L1}$ to $r$, which is also the time it takes the resonance to move from $r_{L1}$ to $r_{L0}$. Hence we have the two equations $\delta t=(r-r_{L1})/v_g$ and $\delta t=(r_{L0}-r_{L1})/v_L$. Combining these two expressions, we find:  $(r-r_{L1})/v_g=(r_{L0}-r_{L1})/v_L$, which allows us to relate the distance $r-r_{L1}$ to the distance from the current resonant location $r-r_{L0}$:
$$r-r_{L0}=r-r_{L1}+r_{L1}-r_{L0}=(1-v_L/v_g)(r-r_{L1})$$
Note that $r-r_{L1}$ corresponds to the distance between the observation point and the resonant radius when the resonant location is fixed. By contrast, $r-r_{L0}$ is the distance between the observed point and the current resonance position. Thus a wave with a moving resonance has its radial profile distorted by a factor of $1-v_L/v_g$. Now, if this is really an $m=-1$ wave, it should be propagating inwards, so $v_g<0$, but the wave also appears to be moving inwards, so $v_L<0$ as well. If $|v_L|>|v_g|$, then $1-v_L/v_g$ will be negative, and the wave will appear on the opposite side of the {\it current} resonance location from what one would expect for a wave with a fixed resonance frequency. 

To clarify what the wave itself would look like in this situation, let us first consider a standard density wave with a fixed resonant radius. In this case, the phase of the wave is given by the following expression:
\begin{equation}
\phi_{tot}=\phi+\phi_r(r)=|m|(\lambda-\Omega_Pt)+\int_{r_L}^rk(r') dr',
\label{pheq1}
\end{equation}
Where $k(r')$ is the radial wavenumber. For waves with $m\ne +1$, $k(r')$ is given by the following asymptotic expression:
\begin{equation}
k(r')\simeq \frac{3M_s}{2\pi\sigma_0 r_L^4}(m-1)(r'-r_L)=\chi(m-1)(r'-r_L)
\end{equation} 
where $M_s$ is Saturn's mass and $\sigma_0$ is the undisturbed surface mass density of the ring.\footnote{Note that $k$ is positive exterior to the resonance for $m>0$ (ILRs) and interior to the resonance for $m<0$ (OLRs).} If we now consider a wave produced by  a moving resonance, we need to replace $r_L$  in the above expressions with either $r_{L0}$ or $r_{L1}$. As mentioned above, the wavenumber of a density wave is determined by how far the wave has propagated from its source region, so the factor of $(r'-r_L)$ in the above expression for $k(r')$ should be replaced with  $(r'-r_{L1})=(r'-r_{L0})/(1-v_L/v_g)$, so in this case the expression for the wavenumber becomes:
\begin{equation}
k(r')\simeq \chi(m-1)\frac{(r'-r_{L0})}{(1-v_L/v_g)}
\end{equation} 
Since $k$ must must be positive, this means that if $|v_L|>|v_g|$ the wave will appear on the opposite side of $r_{L0}$ and the radial trends in the pattern's wavelength will be reversed, as we observe. On the other hand, when we integrate the wavenumber to obtain the phase, we should choose $r_{L0}$ as the lower limit of integration, because we are considering the appearance of the wave at a single moment in time. Hence we expect that the appropriate generalization of Equation~\ref{pheq1} to be
\begin{equation}
\phi_{tot}=|m|(\lambda-\Omega_Pt)+\frac{\chi}{2}(m-1)\frac{(r-r_{L0})^2}{(1-v_L/v_g)}
\end{equation}
If we observe the ring at one time, then a line of constant phase is described by the following equation:
\begin{equation}
\frac{d\lambda}{dr}=-\frac{\chi(m-1)}{|m|}\frac{(r-r_{L0})}{(1-v_L/v_g)}.
\end{equation}
Note that if $|v_L|>|v_g|$, both $1-v_L/v_g$ and $r-r_{L0}$ change sign, which leaves the slope of this curve unchanged. Hence the wave remains a trailing spiral regardless of whether $|v_L|>|v_g|$ or not, which is consistent with our observations (see Figure~\ref{W85.67phase})

We can therefore accommodate the pattern speed and the morphology of the wave so long as the resonance moves through the ring faster than the group velocity, which effectively turns the wave ``inside-out". Furthermore, this slow drift in the resonant radius induces slight phase shifts that slightly modify the wave's apparent pattern speed. This probably explains why the best-fit pattern speed in Figure~\ref{W85.67res} is slightly slower than expected and the phase difference residuals have a rather large scatter.

At present, we have no clear explanation for why the resonant radius would be migrating in the first place, but we strongly suspect this feature is driven by some structure inside Saturn. Note the pattern speed of this structure is 2430$^\circ$/day, which is roughly three times Saturn's rotation period. Since the pattern speed is much faster than the planet's internal rotation rate, this structure cannot be a persistent anomaly carried around by Saturn's winds. It might represent some sort of normal-model oscillation, but sectoral ($\ell=m$) modes with $m=1$ are forbidden (because they would entail a displacement of the planet's center of mass). Modes such as [$\ell=3, m=1$] are possible and could potentially drive $m=-1$ waves in the rings, but these resonances were not predicted to occur at W85.67's location \citep{MarleyPorco93, Marley14}. Another intriguing coincidence is that in a reference frame rotating with Saturn, this pattern would be moving prograde at roughly twice the planet's rotation rate. This is the maximum rotation rate allowed for inertial waves inside a fluid planet \citep{Wu05}. Even though inertial waves do not generate strong gravitational perturbations, this  coincidence might suggest some  connection with those sorts of oscillations (we thank J. Fuller for pointing this out). In any case, the steady change in the oscillation frequency would imply secular evolution of something in Saturn's interior.

\section*{Summary}

Table~\ref{sumtab}  provides the $m$-numbers and pattern speeds of all the previously-unidentified waves securely identified by \citet{HN13} and this work, along with the corresponding  estimates of the ring's surface mass density and opacity for those features  (based on fits by Baillie {\it et al.} 2011). We may summarize the results of this analysis as follows:
\begin{itemize}
\item Wave W83.63 appears to be an $m=-10$ wave, which could be generated by a resonance with an $m=10$ fundamental sectoral normal mode in the planet. 
\item If W83.63 is generated by an $m=10$ planetary normal mode, then that mode must have a larger amplitude than the modes with $m=5-9$, which do not appear to generate waves of comparable strength.
\item The identification of wave W81.02 is still uncertain. It could be $m=-11$ or $m=-5$. The former result seems more consistent with the observations, but the wave's location could favor the latter option.
\item There are no less than five $m=+3$ waves with pattern speeds between 807$^\circ$/day and $834^\circ$/day. Since these pattern speeds are close to the planet's rotation rate, they are probably driven by 3:2 tesseral resonances with persistent gravitational anomalies in the planet.
\item The surface mass densities of the plateaux derived from the $m=+3$ waves are comparable to or less than the surface mass densities of the background ring  derived from inward-propagating waves. The optical depth enhancement in the plateaux therefore must be due to a change in the internal mass densities or size distribution of the ring particles, rather than to the amount of ring material per unit area.
\item The W85.67 wave appears to be an $m=-1$ pattern that is drifting inwards through the rings at a rate of about 0.8 km/year. The high pattern speed of this structure suggests it is generated by a resonance with some dynamic structure inside Saturn, but it remains unclear what  sort of planetary perturbation produces this feature, or why the resonance location is moving.
\end{itemize}

\section*{Acknowledgements}

We wish to thank NASA, the Cassini Project and especially the VIMS Team for providing the data used in this analysis. We also thank R.G.  French for providing refined geometry information for the occultation data. We had useful conservations K. Bailli\'e and J. Fuller, and M. Marley provided helpful comments on the first draft of this paper.


\end{document}